\documentclass[a4paper,11pt]{article}
\pdfoutput=1
\usepackage{jheppub}
\usepackage{bm}

\newcommand{\D}{\mathrm{d}}

\renewcommand{\O}{\mathcal{O}}
\newcommand{\W}{\mathcal{W}}
\newcommand{\<}{\langle}
\renewcommand{\>}{\rangle}
\newcommand{\bs}[1]{\boldsymbol{#1}}
\newcommand{\nn}{\nonumber}
\newcommand{\lla}{\langle \! \langle}
\newcommand{\rra}{\rangle \! \rangle}

\newcommand{\reg}[1]{\hat{#1}}

\newcommand{\dreg}{\hat{d}}
\newcommand{\Dreg}{\hat{\Delta}}

\newcommand{\fr}{\varphi_L}
\newcommand{\Psrc}{\Psi}
\newcommand{\Fsrc}{F}
\newcommand{\e}{{\rm e}}

\DeclareMathOperator{\arctanh}{artanh}

\DeclareMathOperator{\arcosh}{arcosh}

\renewcommand{\(}{\left(}
\renewcommand{\)}{\right)}
\newcommand{\be}{\begin{equation}}
\newcommand{\ee}{\end{equation}}

\newcommand{\secref}[1]{Section \ref{#1}}

\title{Dimensional regularization for holographic RG flows}
\author{Adam Bzowski}
\author{and Marjorie Schillo}
\affiliation{Department of Physics and Astronomy, Uppsala University, Box 516, SE-75120, Uppsala, Sweden}

\emailAdd{adam.bzowski@physics.uu.se}
\emailAdd{marjorie.schillo@physics.uu.se}

\begin{document}

\abstract{
In this work, we present a holographic renormalization scheme for asymptotically anti-de Sitter spacetimes in which the dual renormalization scheme of the boundary field theory is dimensional regularization.  This constitutes a new level of precision in the holographic dictionary and paves the way for the exact matching of scheme dependent quantities, such as holographic beta functions, with field theory computations.  Furthermore, the renormalization procedure identifies a local source field which satisfies the equations of motion along renormalization group flows, resolving a long-standing puzzle regarding the  Wilsonian coupling in holography.  This identification of the source field also provides new insight into field theories deformed by marginal operators, which have been traditionally difficult to analyze due to altered bulk asymptotics. Finally, we demonstrate a new relation equating the analyticity of the holographic beta function to the absence of conformal anomalies, and conjecture that the conformal anomaly should vanish in the UV for all holographic constructions.
}

\maketitle

\section{Introduction}

Just as the predictive computability of any quantum field theory relies on the renormalization of divergences and coupling constants, our ability to compute meaningful quantities using AdS/CFT relies on holographic renormalization.  The procedure for the regularization and renormalization of ultraviolet (UV), or short distance, divergences in quantum field theory (QFT) lies at heart of any QFT textbook.  The AdS/CFT correspondence is a strong/weak coupling duality, so it maps these UV divergences of the boundary field theory into  infrared (IR), or infinite volume divergences of the bulk physics.  The problem of the regularization and renormalization of  bulk divergences was  solved by methods of \emph{holographic renormalization}, \cite{deHaro:2000vlm,Bianchi:2001de,Bianchi:2001kw}. 

Holographic renormalization can be applied in any bulk spacetime that is asymptotically locally AdS.  Thus, using asymptotically AdS domain walls, one can study the renormalization group (RG) flow away from a deformed CFT, where the energy scale of the boundary field theory is related to the radial position in the bulk geometry.  Since the boundary value of a bulk scalar field corresponds to the value of the coupling, or \emph{source}, in the UV CFT, it was argued in \cite{Akhmedov:1998vf} that the value of the field at a given radial position should correspond to the value of the coupling at this scale. Furthermore, in \cite{Girardello:1998pd, Freedman:1999gp} the concept of a holographic beta function was introduced to quantify the running coupling with respect to the radial rescalings. Using the Hamilton-Jacobi formalism to obtain first order equations of motion, \cite{deBoer:1999tgo} identified the equations governing radial bulk evolution with the QFT RG equation. 

One can try to identify how  field theoretic information about the running coupling corresponds to the localized behavior of bulk fields. This is a natural question in the context of Wilsonian RG flow, where changes in the effective action are measured as  UV degrees of freedom are integrated out. The initial proposal \cite{Balasubramanian:1999jd} and later refinements \cite{Heemskerk:2010hk,Faulkner:2010jy} for a holographic construction of Wilsonian field theory defined an effective theory in which UV degrees of freedom are integrated out by performing  the bulk path integral in the region exterior to some radial slice. The value of the Wilsonian coupling at the inverse-energy scale, $L$, defined in this way does not necessarily obey bulk equations of motion and so cannot be identified with the value of the bulk field at a radial position $L$.

Additionally, in the Wilsonian approach, the value of the bulk field at $L$ is generally a complicated functional of both the leading near-boundary field behavior, given by the `source coefficient,' and the sub-leading  `vev coefficient' which corresponds to the vacuum expectation value of the dual boundary operator. Hence, the Wilsonian coupling becomes a non-local function of the CFT source, something rarely observed in QFT.   In \cite{Balasubramanian:2012hb} it was shown that there exists a special `maximum subtraction' scheme, where a running coupling, $\fr$, obeys bulk equations of motion with $L$ identified as a radial variable, but such a solution is not necessarily regular in the IR.

In this paper we resolve these issues -- that the running coupling should be local and satisfy bulk equations of motion --  by developing a renormalization procedure that corresponds to a known, physical, QFT renormalization scheme. This is an extension of the scheme proposed in \cite{Bzowski:2016kni}, where the bulk renormalization procedure corresponds to dimensional regularization in the QFT. The special renormalization scheme found in \cite{Balasubramanian:2012hb} can be shown to be related to the dimensional regularization scheme developed here. We will construct a renormalized coupling constant, $\fr$, at a given scale that satisfies the equations of motion where $L$ is identified with the radial variable. This allows the interpretation that the running coupling constant, with an on-shell renormalization condition, can be identified with a bulk \emph{source field}, $\Psrc_L(z)$, via $\kappa\Psrc_L(L) = L^{d - \Delta} \fr$, where $\kappa$ is the gravitational coupling. 

This  provides a novel understanding of the renormalized coupling constant from the point of view of  bulk physics. If $\Phi$ denotes the `usual' bulk field dual to a given conformal primary operator $\O$ and satisfying Dirichlet boundary conditions, then $\Psrc_L$ will obey Neumann conditions, with a vanishing vev coefficient at the boundary. We can think about $\Psrc_L$ as a bulk field dual to the boundary source, $\fr$, in the sense that $\Phi$ is dual to $\O$.

Furthermore, we will find that the  beta function associated to the dimensionless coupling, $g_L = L^{d-\Delta} \fr$, is  proportional to the derivative of the prepotential, $W'$, as opposed to the usual holographic beta function $W'/W$  \cite{Girardello:1998pd, Freedman:1999gp, deBoer:1999tgo}.  This occurs because the beta function is scheme-dependent, and while in previous work the boundary renormalization scheme was unknown, here we can make a more precise entry into the holographic dicitonary.  The beta functions computed using the source renormalization procedure presented here correspond to QFT beta functions in dimensional regularization.

In addition to added dictionary precision, our methods represent progess in that they can be applied also to CFTs with irrelevant or marginal deformations.  The application of holographic renormalization to irrelevant deformations is usually regarded as intractable due to the lack of appropriate boundary conditions. In this paper we will present an example of a holographic RG flow driven by an irrelevant operator based on \cite{Berg:2001ty}. We show how dimensional methods deal with the asymptotics and uniquely determine source and vev coefficients in the near-boundary expansion.

Finally, we are able to addresses recent questions raised in the case of a bulk scalar field dual to a (classically) marginal operator, where  a tower of logarithmic divergences in the near-boundary field expansion spoils AdS boundary conditions. The prime example of this behavior is \cite{Klebanov:2000hb}. As in perturbative QFT, in order to identify the source, one needs to carry out a renormalization procedure. This problem was initially analyzed in  \cite{Aharony:2005zr} and further studies  \cite{Borodatchenkova:2008fw,Haack:2010zz,Muck:2010uy,Nakayama:2013fha} contain partial results. More recently, standard holographic methods have been applied to nearly marginal flows \cite{Bourdier:2013axa,Kiritsis:2014kua}.
A physical interpretation of all these results, however, is still lacking, since their QFT schemes  remain obscure. Despite recent efforts in the general analysis of holographic renormalization schemes in \cite{Lizana:2015hqb,Sathiapalan:2017frk}, it is difficult to identify a specific scheme. The identification of both the boundary renormalization scheme and the running coupling in terms of bulk fields presented here provides a comprehensive renormalization procedure for the holographic theories deformed by a marginal operator.

\section{Holographic set-up}

The original application of the AdS/CFT correspondence, and the one we will pursue here, is to use a weakly coupled gravitational system in an asymptotically AdS spacetime to define a dual QFT non-perturbatively.  The weakly coupled gravitational system can be described by the Einstein-Hilbert action coupled matter; we will exclusively focus on scalar matter.  Then, the bulk Euclidean action reads
\begin{align} \label{preS_AdS}
S_E & = \int \D^{d} \bs{x} \D r \sqrt{g} \left( - \frac{R}{2 \kappa^2} + \frac{1}{2} \partial_\mu \Phi \partial^\mu \Phi + V(\Phi) \right)\,.
\end{align}
where the scalar potential $V(\Phi)$ has a regular Taylor expansion around $\Phi = 0$:
\begin{equation} \label{V}
V(\Phi) = - \frac{d(d-1)}{2 l^2 \kappa^2} + \frac{\Delta (\Delta - d)}{2 l^2} \Phi^2 + \sum_{n=3}^{\infty} \frac{\lambda_n}{n l^2 \kappa^2} (\kappa \Phi)^n\,.
\end{equation}
The $(d+1)$-dimensional gravitational coupling, $\kappa$, is related to the reduced Planck mass by $\kappa^{-2} = M_{Pl}^{d-1}$.  We will often find it useful to work with the dimensionless combinations $\kappa \Phi$. The scalar mass is given by $m^2l^2 = \Delta(\Delta-d)$, where $\Delta$ is the conformal weight of the dual scalar operator in the CFT and $l$ is the AdS radius.

The requirement that the bulk geometry is asymptotically AdS implies that metric admits the Fefferman-Graham gauge, in Poincare\'{e} coordinates this reads
\begin{align}
\D s^2 & = \D r^2 + \gamma_{ij} \D x^i \D x^j && \gamma_{ij} = e^{2 r/l} \gamma_{(0)ij} + \text{sub-leading in } r \rightarrow \infty \text{ limit}\,.
\end{align}
The equation of motion for the scalar field following from \eqref{preS_AdS} is given by the  Klein-Gordon equation,
\begin{equation} \label{eqofmo_AdSa}
\left( \partial_r^2 - K \partial_r + \Box_{\gamma} \right) \Phi(r, \bs{x}) - V'(\Phi(r, \bs{x})) = 0\,,
\end{equation}
where $K$ is the trace of the extrinsic curvature of the hypersurface defined with respect to the unit normal $\partial_r$, and $\Box_{\gamma}$ is defined with respect to $\gamma_{ij}$.
In an asymptotically AdS spacetime the extrinsic curvature has the expansion $K = -d/l + \ldots$, where the omitted terms vanish at the boundary, $r\to \infty$. The second order differential equation \eqref{eqofmo_AdSa} has two independent solutions whose leading behaviors are proportional to $e^{-(d-\Delta)r/l}$ and $e^{-\Delta r/l}$.

In what follows it will be  convenient to use a different coordinate defined as
\begin{equation}
z = l e^{-r/l}\,,
\end{equation}
where the conformal boundary now lies at $z=0$. The radial or near-boundary expansion of the scalar field $\Phi$ is given by:
\begin{equation} \label{Phi_exp}
\kappa \Phi = \big( \phi_{(d - \Delta)} z^{d - \Delta} + \ldots\big) + \(\phi_{(\Delta)} z^{\Delta} + \ldots\)\,,
\end{equation}
where the omitted terms are necessarily sub-leading only within each set of parenthesis.  For $d/2<\Delta<d$ the leading behavior of the scalar field is given by  $\kappa \Phi \sim  \phi_{(d - \Delta)} z^{d - \Delta}$.

The AdS/CFT correspondence states that there exists a one-to-one  map between single trace conformal primaries in the boundary CFT (which is the UV fixed point of the boundary QFT) and bulk fields in the gravity dual. The generating functional of the dual QFT, $\W$, is given by the bulk on-shell action, $\W[\phi_{(d - \Delta)}] = - S_{\text{on-shell}}[\phi_{(d - \Delta)}]$. Where $\phi_{(d - \Delta)}$, the asymptotic boundary value of the bulk field $\Phi$, is identified as the source of the corresponding operator, $\O_\Delta$, on the field theory side.

In order to derive QFT correlation functions, we must ensure two conditions are satisfied. First, the asymptotic boundary value problem must be well-posed; this requires the addition of the Gibbons-Hawking-York boundary term. Second, on-shell bulk action should be well defined; this requires \emph{holographic renormalization} to regulate divergences in the on-shell action, \emph{e.g.} due to the infinite volume of AdS. 

\subsection{Traditional holographic renormalization}
The standard procedure to extract finite quantities from the divergent supergravity action is holographic renormalization \cite{deHaro:2000vlm,Bianchi:2001kw,Bianchi:2001de}. In this approach one imposes a cut-off surface at some  $z=\delta > 0$ and adds suitable, local, bulk-covariant counterterms supported on this surface. The counterterms are constructed such that after the solution to the equation of motion is substituted, a finite $\delta \rightarrow 0$ limit exists. Including the Gibbons-Hawking-York boundary term and the first two counterterms, which subtract the volume divergences, the bulk action \eqref{preS_AdS}  reads
\begin{align} \label{S_AdS}
S & = \lim_{\delta \rightarrow 0} \left[ \int \D^{d} \bs{x} \int_\delta \D z \sqrt{g} \left( - \frac{R}{2 \kappa^2} + \frac{1}{2} \partial_\mu \Phi \partial^\mu \Phi + V(\Phi) \right)  + \frac{1}{\kappa^2} \int \D^d \bs{x} \sqrt{\gamma} K \right. \nn\\
& \qquad\qquad \left. + \int \D^d \bs{x} \sqrt{\gamma} \left( \frac{d - 1}{l \kappa^2} + \frac{d - \Delta}{2 l} \Phi^2 \right) \right],
\end{align}
where $\gamma$ is the metric induced on the $z = \delta$ cut-off surface. With these two counterterms included, the on-shell action is finite provided the dimensions $d$ and $\Delta$ satisfy\footnote{In the case of a flat boundary, $\gamma_{(0)ij}=\delta_{ij}$, the restriction on $d$ can be removed.}
\begin{equation} \label{Range}
\frac{d}{2} < \Delta < \min \left( \frac{d}{2} + 1, \frac{2 d}{3} \right) \quad\text{and}\quad  0 < d < 2\,,
\end{equation}
otherwise, additional counterterms will be required.

As with the first two counterterms in the second line of \eqref{S_AdS}, most counterterms are uniquely fixed. However, in special cases, counterterms introducing scheme-dependence appear. These terms are related to the emergence of secular terms in near-boundary expansions, \eqref{Phi_exp}, which take logarithmic form: $\log(z\mu)$, where $\mu$,  the renormalization scale, must be introduced on dimensional grounds.   We will refer to counterterms that arise due to secular terms in the near-boundary expansion as \emph{secular counterterms}, and other counterterms \emph{canonical counterterms}. The scheme-dependence of the secular counterterms arises precisely in the freedom to redefine $\mu$. For convenience, we will often use the inverse renormalization scale, $\mu^{-1}=L$.
For example, the usual counterterm action will contain terms such as \cite{Skenderis:2002wp,Papadimitriou:2004ap}:
\be \label{phiboxphiCT}
S_{ct} \propto  \mu^{-(d-2\Delta + 2 k)} \int \D^dx \sqrt{\gamma} \Phi \Box^k \Phi\,,
\ee
where $k$ is a non-negative integer. In the special cases $\Delta = d/2+k$, the $\mu$ dependence appears to vanish, however coefficients of these terms will include $\log(z\mu)$, indicating a secular term proportional to $z^\Delta \log(z\mu)$ in the near-boundary expansion \eqref{Phi_exp}.

\subsection{Holographic dimensional renormalization}
Here, we will develop an alternate approach to traditional holographic renormalization, dubbed \emph{holographic dimensional renormalization}.  This method of holographic renormalization was first introduced in \cite{Bzowski:2016kni}, and in this work it will be expanded to the case of RG flows and domain wall spacetimes.  This procedure makes use of the observations that \eqref{Range} defines an open and non-empty subset of the parameter space $(d, \Delta)$, and the bulk equations of motion are analytic.  These observations allow one to analytically continue $d$ and $\Delta$ as in the familiar QFT dimensional regularization:
\begin{equation} \label{dimreg}
 \dreg = d + u \epsilon \quad \text{and} \qquad \Dreg = \Delta + v \epsilon\,.
\end{equation}
The constants $u$ and $v$ indicate a direction of the infinitesimal shift in the space of dimensions $(d, \Delta)$ and $\epsilon$ is the regulator.\footnote{In the context of textbook QFT the shift in dimensions is such that in momentum space bare propagators of fundamental fields retain their canonical form, such as $1/k^2$ for a massless scalar propagator. }
After a correlation function is evaluated for a generic $\dreg$ and $\Dreg$, one can continue it away from the parameter space \eqref{Range}. If the result is well-defined, it represents the unique correlation function.  Analyticity in $d$ and $\Delta$ was demonstrated in the context of 3-point functions in a CFT  in \cite{Bzowski:2015yxv} and in the context of scalar fields in holographic spacetimes in \cite{Bzowski:2016kni}.

The use of analytic continuation of spacetime and operator dimension reduces the number of counterterms needed to renormalize the on-shell action.  To begin with, other than the first two volume divergence counterterms in \eqref{S_AdS}, all canonical counterterms are absent; naive divergences disappear when the on-shell action is defined for general $\dreg$ and $\Dreg$.  This is natural from the boundary field theory perspective where the $z\rightarrow 0$ limit corresponds to flowing to the CFT at the UV fixed point of the boundary QFT.  Classically, in the CFT there can be no explicit scale dependence, so the only allowable counterterms are secular, where the scale appears only logarithmically. For example, of the counterterms in \eqref{phiboxphiCT}, only those where $\Dreg = \dreg/2 + k +O(\epsilon)$ will be needed in holographic dimensional renormalization.  Thus, in the $\epsilon \rightarrow 0$ limit one recovers the form or an appropriate CFT counterterm.  This one-to-one correspondence between bulk and CFT counterterms is the first indication that holographic dimensional renormalization corresponds to a known, well-defined, field theory renormalization scheme.

In this paper we are interested in the analysis of holographic RG flows using holographic dimensional renormalization. In the context of perturbative QFT, RG flows are induced by the existence of non-zero beta functions. These in turn, emerge through the renormalization of coupling constants. For the classically marginal scalar operator $\O$ studied most often in textbook QFT, the original `bare' coupling $\phi_0$ is renormalized by the addition of (usually infinite) counterterms, resulting in a renormalized coupling, $\fr$. The general form for such QFT counterterms is:
\begin{equation} \label{ftct}
S_{\text{ct}} = \int \D^{\reg{d}} \bs{x} \: \fr Z[\fr L^{\epsilon}] \O\,,
\end{equation}
where we use $L = \mu^{-1}$ as the inverse-energy scale. The renormalized source $\fr$ is implicitly scale-dependent in such a way that the bare source $\phi_{0} = \fr Z[\fr L^{\epsilon}]$ remains scale-independent. The renormalization factor, $Z$, depends on the combination $g_L=\fr L^{\epsilon}$ which we identify as the renormalized dimensionless coupling.
The generating functional of the renormalized theory - provided no other divergences are present - reads,
\begin{equation} \label{genfun}
\W[\fr] = \lim_{\epsilon \rightarrow 0} \< \exp \left( - \int \D^{\reg{d}} \bs{x} \: \fr Z[\fr L^{\epsilon}] \O \right) \>_{\text{reg}}\,,
\end{equation}
where $\< \ \cdot\  \>_{\text{reg}}$ denotes the connected correlation function in the regulated theory.

In this way, one can see that the beta function is induced by source redefinition. Since the bare source, $\phi_0$, remains scale-free, we can calculate the beta function, $\beta_g$, for the dimensionless coupling $g_L = L^\epsilon \fr$ by noting that the total derivative of $\phi_0(L,g_L)$ with respect to $L$ vanishes,
\begin{equation} \label{betadef}
\beta_g(g_L) = \mu \frac{\D g_L}{\D \mu} = - L \frac{\D g_L}{\D L} = - \epsilon \phi_0(g_L) \left( \frac{\partial \phi_0}{\partial g_L} \right)^{-1}\,.
\end{equation}

We will follow a parallel course in holographic dimensional renormalization, focusing on source renormalization to cancel divergences and induce a holographic beta function.  In \cite{Bzowski:2016kni} it was shown that source renormalization (accompanied by certain secular counterterms) removes the divergences from correlation functions in holographic theories. In the holographic set-up, the bare source is identified with the coefficient $\phi_{(d - \Delta)}$ in \eqref{Phi_exp}.\footnote{This is strictly only true for the standard, Dirichlet, boundary conditions. We will briefly comment on mixed boundary conditions arising from multi-trace deformations in Section \ref{sec:further}.}  The process of source renormalization, which will be outlined in-depth in \secref{marginaldimreg}, will remove additional divergences from the on-shell action. These divergences are related to the emergence of certain secular logarithmic terms in the near-boundary expansion of the bulk field.  Source redefinition is equivalent to adding counterterms of the form \eqref{ftct} and will analogously lead to a beta function as in \eqref{betadef}.

\section{Dimensional renormalization for marginal operators} \label{marginaldimreg}

We will use source redefinition, following QFT intuition, to renormalize divergences which arise when a descendent of a source (\emph{i.e.} one of the omitted terms in the first set of parenthesis in \eqref{Phi_exp}) has the same scaling dimension as another scalar field's source or its descendent. However, the secular counterterms which arise when a descendent of a source has the same scaling dimension as a vev term (\emph{i.e.} $\phi_{(\Delta)}$ or one of its descendants) are still needed in the counterterm action.\footnote{Additionally, for irrelevant deformations counterterms containing canonical momenta may be necessary \cite{vanRees:2011fr,vanRees:2011ir}.} This latter case is what gives rise to the counterterms containing sources only. These counterterms are constructed using only bulk fields and boundary momenta, (\emph{i.e.} not depending on radial derivatives or canonical momenta,) and will induce conformal anomalies. An example of such a secular counterterm containing two bulk fields has the form \eqref{phiboxphiCT} with $\Delta = d/2 +k$. The fact that these secular counterterms remain after source redefinition should not be surprising since in QFT the emergence of beta functions does not preclude anomalies.

The process of source redefinition will provide a solution to the longstanding confusion surrounding the application of holographic renormalization to the case of a marginal deformation by an operator with $\Delta =d$.  In this case the expansion \eqref{Phi_exp} exhibits an infinite tower of secular terms:
\begin{equation} \label{Phi_special}
\kappa \Phi = \psi_{(0)} + \psi_{(1)} \log z + \psi_{(2)} \log^{2} z + \ldots\,.
\end{equation}
These terms spoil boundary asymptotics, due to the lack of a  $z \rightarrow 0$ limit, making the identification of the source problematic. Applications of the general renormalization methods of \cite{Chen:1995ena} help to make progress in a rigid AdS background \cite{Nakayama:2013fha}, but the identification of the source remains unclear. Furthermore, when coupled to gravity, the AdS boundary conditions are spoiled by logarithmic terms as well, \cite{Klebanov:2000hb,Borodatchenkova:2008fw,Haack:2010zz,Muck:2010uy,Papadimitriou:2011qb}.

In the dimensional renormalization approach the secular terms \eqref{Phi_special} emerge in the $\epsilon \rightarrow 0$ limit of the bulk field expansion \eqref{Phi_exp}, with dimensions shifted according to \eqref{dimreg}, as 
\begin{equation} \label{ex1_Phi_src}
\kappa \reg{\Phi} =  \phi_{(w\epsilon)} z^{w\epsilon} +  \phi_{(2 w\epsilon)} z^{2 w\epsilon} +  \phi_{(3 w\epsilon)} z^{3w \epsilon} + \ldots\,,
\end{equation}
where we define $w=u-v$.  We also introduce the notation that a `hat' indicates a regulated quantity, \emph{i.e.} one that depends on $\epsilon$. This is precisely the case where source renormalization removes the need for secular counterterms: an infinite tower of descendants of the source, $\phi_{(w\epsilon)}$, all collapse to have the same scaling dimension.  Since the equations of motion and their solutions are analytic in the bulk, this expansion must converge to \eqref{Phi_special} at any bulk point when $\epsilon \rightarrow 0$. If all coefficients $\phi_{(nw \epsilon)}$ are finite in such limit, the expansion \eqref{Phi_special} would contain only a single non-vanishing term, $\psi_{(0)}$. Hence, the emergence of the logarithmic terms in \eqref{Phi_special} requires that the coefficients $\phi_{(n w\epsilon)}$ are divergent in the $\epsilon \rightarrow 0$ limit. However, the existence of the limit imposes constraints on these divergences. From the point of view of the dual QFT these constraints are equivalent to renormalizability, or the existence of a finite beta function.

One of the main advantages of the dimensional renormalization method is that the problem of marginal deformations can be tackled directly and will uniquely identify the renormalized source and the beta function. Particularly this will allow us to make progress on the long-standing issue of computing correlation functions for the confining gauge theory dual to the Klebanov-Strassler background \cite{Klebanov:2000hb}.  This new method for holographic computations in backgrounds that violate the asymptotic AdS will allow us to advance, building upon prior work in this direction \cite{Papadopoulos:2000gj, Aharony:2005zr, Berg:2005pd}.  This is both complex and subtle and therefore will be treated in a separate work \cite{future}.

We will begin with a presentation of the general procedure for the renormalization of a  scalar source and the resulting beta function. To actualize what may seem a rather abstruse general prescription, we will immediately apply it to the simplest possible example: a scalar on a rigid AdS background.  Next, we will demonstrate how the procedure works in the case of dynamical gravity, making contact with known holographic domain wall results.  This will allow us to identify the scheme corresponding to dimensional regularization on the gravity side of the correspondence. Then, we present an example to demonstrate that holographic dimensional renormalization extends to define  beta functions to all-orders in perturbation theory. Finally, we discuss the effect of holographic dimensional renormalization on the computation of correlation functions, again using the simplest example of a scalar on rigid AdS.

\subsection{General procedure} \label{genprod}

We begin the process of renormalizing the source by solving the equations of motion of the regulated theory order-by-order in a near-boundary expansion.  By `regulated theory' we mean shifting $d$ and $\Delta$ according to \eqref{dimreg}.  For a marginal operator, $\Dreg = d + v \epsilon$, one must solve the Klein-Gordon equation \eqref{eqofmo_AdSa} with the regulated potential and extrinsic curvature:  
\be
\reg{V}(\reg{\Phi})=-{\dreg(\dreg-1) \over 2l^2\kappa^2}-{\epsilon \Dreg \over 2 l^2}\reg{\Phi}^2 + O(\reg{\Phi}^3), \quad \text{and} \quad \hat{K} = -\dreg/l+\cdots\,.
\ee
Here and henceforth we choose $w=u-v=1$ to simplify notation.  Note that the choice of $u$ and $v$ may appear in some scheme-dependent quantities, but by assigning dimensions to $\epsilon$ and inverse dimensions to $u,v,w$, one can always restore the $w$ dependence. In the renormalized theory scheme-dependent terms containing $u,v,w$ may appear in correlation functions. Similarly to textbook QFT, the scheme-dependence can be absorbed into the scale-dependence of the correlators, see \secref{correlationfunctions}.

From this we find that the near-boundary expansion of the regulated field $\reg{\Phi}$ has the form \eqref{ex1_Phi_src} with all coefficients $\phi_{(n \epsilon)}$ for $n \geq 2$ determined in terms of $\phi_{(\epsilon)}$. As anticipated in the text below \eqref{ex1_Phi_src}, the higher order terms $\phi_{(n \epsilon)}$ typically diverge when $\epsilon \rightarrow 0$. By solving equations of motion order-by-order in the source, $\phi_{(\epsilon)}$, near  $z = 0$ one finds the divergent coefficients, $c_{ii}$, in the following expansion:
\begin{equation} \label{ex1_Phi_src1}
\kappa \reg{\Phi} =  \phi_{(\epsilon)} z^{\epsilon} + c_{22}  \phi_{(\epsilon)}^2 z^{2 \epsilon} + c_{33}  \phi_{(\epsilon)}^3 z^{3 \epsilon} + O(\phi_{(\epsilon)}^4, z^2)\,.
\end{equation}

In order to cure these divergences we will introduce the renormalized  source, $\fr$, via a redefinition of the bare source, $\phi_{(\epsilon)}$
\begin{equation} \label{renZ}
\phi_{(\epsilon)} = \fr Z[\fr L^{\epsilon}] = \fr \sum_{n=0}^{\infty} Z_n \fr^n L^{n \epsilon}\,,
\end{equation}
with $Z_0 = 1$. Here, $L$ has been introduced on dimensional grounds and serves the purpose of the inverse renormalization scale, $\mu^{-1}$. Additionally, the renormalized source $\fr$ depends implicitly on scale $L$ in such a way that the bare source $\phi_{(\epsilon)}$ remains scale-independent -- in general, a subscript $L$ will indicate dependence on the renormalization scale. We will choose the coefficients $Z_n$ in order to preserve a finite limit in \eqref{ex1_Phi_src1}. 

Since the bulk equations of motion for the scalar field are second order, picking specific boundary conditions and enforcing bulk regularity will induce non-local dependences between the `source coefficient' $\phi_{(\dreg - \Dreg)}$ and the `vev coefficient' $\phi_{(\Dreg)}$. The regulated field, $\reg{\Phi}$, may represent any of these solutions, with arbitrary vev coefficient. The renormalization of the sources, however, deals with source redefinition only, and hence we drop all terms which depend on the vev coefficient; in the marginal case this means dropping terms of order $z^\alpha$ where $\alpha$ remains finite in the $\epsilon \to 0$ limit. In this way we define the regulated \emph{source field} $\reg{\Psrc}$,
\begin{align} \label{Phi_src}
\kappa \reg{\Psrc} = \left. \kappa \reg{\Phi} \right|_{z^{O(\epsilon)}} & = \sum_{n=1}^\infty \phi_{(n \epsilon)}(\phi_{(\epsilon)}) z^{n \epsilon} \nn\\
& = \sum_{n=1}^\infty c_{nn}(\epsilon) \phi_{(\epsilon)}^n z^{n \epsilon}\,,
\end{align}
which remains the same regardless of boundary conditions imposed on the physical bulk field $\reg{\Phi}$. 

Inserting \eqref{renZ} into \eqref{Phi_src} produces the regulated source field as a function of the renormalized source,
\begin{equation} \label{Phi_src_reg}
\begin{array}{rllll}
\kappa \reg{\Psrc}_L = z^{\epsilon} \left[ \right. & \fr & + Z_2 L^{\epsilon} \fr^2 & + Z_3 L^{2 \epsilon} \fr^3 &\left. + \ldots \right] \\
+ z^{2 \epsilon} \left[ \right. & & + c_{22} \fr^2 & + c_{23} L^{\epsilon} \fr^3 &\left. + \ldots \right] \\
+ z^{3 \epsilon} \left[ \right. & & & + c_{33} \fr^3 &\left. + \ldots \right] \\
\ldots\,. &&&&
\end{array}
\end{equation}
Coefficients $c_{nn}$ remain the same as dictated by the equations of motion \eqref{ex1_Phi_src1}, however `cross-terms' arising from the substitution of \eqref{renZ} will generate non-diagonal $c_{ij}$ coefficients at each order in $z^\epsilon$.  One can easily check that each $c_{ij}$ is determined in terms of $c_{i'j'}$ with $i' < i$ and $j' < j$. Hence one can choose coefficients $Z_n$ such that the sum of terms in each column is finite when $\epsilon \rightarrow 0$. Simply, one chooses
\begin{equation} \label{Zn}
Z_n = - \sum_{j=2}^n c_{jn} + Z_n^{(0)}\,,
\end{equation}
where $Z_n^{(0)}$ is an $\epsilon$-independent constant.  The choice of $Z_n^{(0)}$ corresponds to scheme-dependence, and we will always choose $Z_n^{(0)}=0$ below.
With this prescription for the $Z_n$, the finite $\epsilon \rightarrow 0$ limit of $\reg{\Psrc}_L$ exists and will be denoted by $\Psrc_L$.  If the theory does not contain any other sources of divergence, such as anomalies, the finite $\epsilon \rightarrow 0$ limit of the full bulk field and the on-shell action exists. The near-boundary expansion takes form \eqref{Phi_special} and an infinite tower of logarithms is present.

By comparison with \eqref{genfun} we see that the QFT bare source $\phi_0$ is identified with the regulated source $\phi_{(\epsilon)}$. The renormalized source is then $\fr$ and the $Z_n$ factors in \eqref{renZ} are the multiplicative renormalization factors of the dual dimensionally regulated QFT. This identification constitutes the only instance of a holographic renormalization procedure where the boundary field theory regularization is known.

While the renormalization procedure is in its essence perturbative, we can provide its definition to all orders in $\phi_{(\epsilon)}$ as follows. First, notice that the source field $\reg{\Psrc}$ in \eqref{Phi_src} depends on the radial variable and the bare source through the combination $z^{\epsilon} \phi_{(\epsilon)}$, \textit{i.e.}, $\kappa \reg{\Psrc} = \Fsrc(z^{\epsilon} \phi_{(\epsilon)})$ for some function $\Fsrc$. We define the field $\reg{\Psrc}_L$ in \eqref{Phi_src_reg}, by redefining $\phi_{(\epsilon)}$ as a function of the renormalized source $\fr$ in such a way that the finite $\epsilon \rightarrow 0$ limit exists. Notice that with all $Z_n^{(0)} = 0$ in \eqref{Phi_src_reg} the procedure is equivalent to
\begin{equation} \label{src_ren}
\kappa \reg{\Psrc} = \Fsrc(z^{\epsilon} \phi_{(\epsilon)}),\, \qquad\qquad L^{\epsilon} \phi_{(\epsilon)} = \Fsrc^{-1}(\fr L^{\epsilon}) = F^{-1}(g_L)\,.
\end{equation}
Indeed, since $Z_n^{(0)} = 0$ in equation \eqref{Zn}, the coefficients $Z_n$ are Taylor coefficients of the expansion of $\Fsrc^{-1}$. In particular
\begin{equation} \label{ren_cond}
\kappa \reg{\Psrc}_L(z = L) = \varphi_L L^\epsilon = g_L\,, \qquad\qquad \kappa \Psrc_L(z = L) = g_L\,,
\end{equation}
where $g_L = \varphi_L L^\epsilon$ is the dimensionless coupling constant.

Finally, we are ready to define the holographic beta function corresponding to dimensional regularization in the boundary field theory.  Again, note that $\phi_{(\epsilon)}$ does not depend on $z$ or $L$ and hence the $z$ derivative of $\kappa \reg{\Psrc}$ matches the $L$ derivative of $g_L$ in \eqref{src_ren},
\begin{equation} \label{betag}
\beta_g = - L \frac{\D g_L}{\D L} = \left. - \kappa z \frac{\D \reg{\Psrc}_L}{\D z} \right|_{z = L}\,.
\end{equation}
This analysis holds for arbitrary values of $\epsilon$ and will therefore immediately generalize to the analysis of relevant flows. In the context of marginal deformations considered here, one can take the $\epsilon \rightarrow 0$ limit and the regulated source field $\reg{\Psrc}_L$ becomes $\Psrc_L$. Thus, this procedure shows that $\kappa \Psi_L$ is \emph{the running coupling} as it satisfies the RG equation \eqref{betag}. Furthermore, with the choice $Z_n^{(0)} = 0$ the running coupling satisfies the normalization condition \eqref{ren_cond}, which is the \emph{on-shell renormalization scheme}. In the on-shell scheme the value of the source field, $\Psi_L$ at the inverse energy scale $z=L$ equals the physical coupling constant $g_L$ at this scale.
  
With the definition \eqref{ren_cond} we can expand the renormalized source field, $\Psrc_L$, matching with the expansion \eqref{Phi_special} where each logarithmic term becomes $\log(z/L)$.  Hence, the leading term $\psi_{(0)}$ can be identified with the source, $g_L$ at scale $L$.  It is the inclusion of this scale dependence which leads to non-trivial beta functions.\footnote{In \cite{Papadimitriou:2011qb} holographic renormalization is carried out for marginal operators, however this scheme explicitly fixes the scale such that $g_L=1$, precluding the calculation of a beta function as in \eqref{betag}.}  
  
By construction, the source field satisfies the bulk equations of motion and does not depend on the vev coefficient, $\phi_{(\Dreg)}$.  These two conditions allow us to identify the dimensional renormalization scheme as the zero-momentum limit of the `maximal subtraction scheme' of \cite{Balasubramanian:2012hb}. Unlike the scheme of \cite{Balasubramanian:2012hb}, the redefinition of the source presented here satisfies the standard QFT expectation that the renormalized source is a local function of the bare source.  This is accomplished by the fact that the source does not depend on the vev and \emph{additionally} that it does not depend explicitly on momentum.  This indicates that while the source field $\Psrc_L$ is a solution of the equations of motion, it is not equal to the full solution $\Phi$ which is used to compute correlation functions. Nonetheless, the redifinition of the source will have an effect of the computation of correlation functions as we will see in \secref{correlationfunctions}.

\subsection{Example: rigid AdS}
In this section we apply the procedure outlined above to the simplest possible case: a massless scalar field on a rigid AdS background with the dynamics governed by a regular potential of the form
\begin{equation}
V(\Phi) = \frac{\lambda_3}{3 l^2 \kappa^2} (\kappa \Phi)^3 + \frac{\lambda_4}{4 l^2 \kappa^2} (\kappa \Phi)^4 + O(\Phi^5)\,.
\end{equation}
 The regulated potential is
\begin{equation}
\reg{V}(\reg{\Phi}) = - \frac{\epsilon(d + v \epsilon)}{2 l^2}   \reg{\Phi}^2 + \frac{\reg{\lambda}_3}{3 l^2 \kappa^2} (\kappa \reg{\Phi})^3 + \frac{\reg{\lambda}_4}{4 l^2 \kappa^2} (\kappa \reg{\Phi})^4 + O(\reg{\Phi}^5)\,.
\end{equation}
In principle, the coefficients $\reg{\lambda}_j$ in the regulated potential can depend on the regulator as well, $\reg{\lambda}_j = \reg{\lambda}_j(\epsilon)$, in such a way that we recover original coefficients in the $\epsilon \rightarrow 0$ limit. Generally, the sub-leading terms in $\epsilon$ will be subleading in the solution and will therefore not affect the beta function.\footnote{However, we will see that this is explicitly not the case for certain relevant deformations in \secref{gppz}.}

The Klein-Gordon equation on the rigid AdS is given by \eqref{eqofmo_AdSa} with $\reg{K} = -\reg{d}/l$. Solving for the first two coefficients in the expansion \eqref{ex1_Phi_src1}, one finds:
\begin{align}
c_{22} & = - \frac{\reg{\lambda}_3}{\epsilon \; (d + (v - 2) \epsilon)}\,, \label{c22} \\
c_{33} & = \frac{\reg{\lambda}_3^2}{\epsilon^2 \; (d + (v - 2) \epsilon) (d + (v - 3) \epsilon)} - \frac{\reg{\lambda}_4}{2 \epsilon \; (d + (v - 3) \epsilon)}\,. \label{c33}
\end{align}
As expected, the coefficients diverge at $\epsilon = 0$. Using the result \eqref{ren_cond} we recognize that the renormalized dimensionless source $g_L = L^\epsilon \fr$ is given by \eqref{ex1_Phi_src1} evaluated at $z=L$ and dropping all terms with vev (and momentum) dependence:
\begin{equation}
g_L = L^\epsilon \fr = L^{\epsilon} \phi_{(\epsilon)} + c_{22} L^{2 \epsilon} \phi_{(\epsilon)}^2 + c_{33} L^{3 \epsilon} \phi_{(\epsilon)}^3 + O(\phi_{(\epsilon)}^4)\,.
\end{equation}
Then, solving for the $Z_n$ is equivalent to inverting the power series for $\phi_{(\epsilon)}$,
\begin{equation} \label{ex1_phi0}
L^\epsilon \phi_{(\epsilon)} = g_L - c_{22} g_L^2  + (2 c_{22}^2 - c_{33} ) g_L^3 + O(g_L^4)\,.
\end{equation}
Now the finite $\epsilon \rightarrow 0$ limit of $\reg{\Psi}_L$ \eqref{Phi_src_reg} exists, order by order in $g_L$, and we find
\begin{align}
& \kappa \Psi_L = \lim_{\epsilon \rightarrow 0} \kappa \reg{\Psi}_L = g_L - \frac{\lambda_3}{d} \log \frac{z}{L} g_L^2 \nn\\
& \qquad\qquad + \left( \frac{\lambda_3^2}{d^2} \log^2 \frac{z}{L} + \frac{2 \lambda_3^2 - \lambda_4 d^2}{d^3} \log \frac{z}{L} \right) g_L^3 + O(g_L^4)\,. \label{renPhi}
\end{align}

To calculate the beta function in the dual QFT directly, we can use equation \eqref{betadef}. By expanding coefficients $c_{nn}$ in $\epsilon$ and keeping leading terms only we find
\begin{align} \label{ex_beta}
\beta_g(g_L) & = - \epsilon g_L + \frac{\lambda_3}{d} g_L^2 + \frac{\lambda_4 d^2 - 2 \lambda_3^2}{d^3} g_L^3 + O(g_L^4, \epsilon)\,.
\end{align}
Here we see that in the $\epsilon \rightarrow 0$ limit all $u,v,w$-dependence vanishes from the beta function. We have also included the customary classical factor $-\epsilon g_L$, which obviously vanishes in the $\epsilon \rightarrow 0$ limit.  On the other hand, the same result can be obtained from \eqref{betag} directly using  $\kappa \Psi_L$ from \eqref{renPhi}. The finiteness of the beta function in a QFT follows from renormalizability of the theory. In the context of holographic theory, this manifests through the existence of the $\epsilon \rightarrow 0$ limit of the regulated solution $\reg{\Psi}_L$.

\subsection{Holographic dimensional renormalization for domain walls} \label{marginalDW}
We now come to the physically interesting case of dynamical gravity.  In this section we will apply holographic dimensional renormalization to the system of a marginal scalar coupled to gravity and governed by the action \eqref{S_AdS} with a regulated potential 
\begin{equation} \label{VDW}
\hat{V}(\Phi) = - \frac{\dreg(\dreg-1)}{2 l^2 \kappa^2} - \frac{\epsilon \Dreg}{2 l^2} \reg{\Phi}^2 + \sum_{n=3}^{\infty} \frac{\reg{\lambda}_n}{n l^2 \kappa^2} (\kappa \reg{\Phi})^n\,.
\end{equation}
The near-boundary expansion for the domain wall metric ansatz is
\begin{equation} \label{greg}
g_{\mu\nu} \D x^\mu \D x^\nu =  \D r^2 + \e^{2 \reg{A}(r)} \left[ \gamma_{(0)ij} + O(\e^{-2r/l}) \right] \D x^i \D x^j\,.
\end{equation}

It is well known that in the homogeneous case, where the scalar field and the metric depend on the radial coordinate only, the system substantially simplifies and the two second order equations of motion for the scalar field, $\Phi$, and scale factor, 
\begin{equation}
a(r)=\e^{A(r)}\,, 
\end{equation}
can be traded for two first order equations, 
\be 
\begin{split}
\partial_r \Phi & = W'(\Phi)\,, \label{eqPhi} \\
\partial_r A & = - \frac{\kappa^2}{d - 1} W(\Phi)\,,
\end{split}
\ee
in addition to a non-linear equation for the prepotential: $W$,
\begin{equation} \label{VOfW}
V = \frac{1}{2} (W')^2 - \frac{d \kappa^2}{2 (d - 1)} W^2\,.
\end{equation}

The addition of dynamical gravity does not present any obstacle; the equations of motion in the regulated theory can be solved perturbatively in a near-boundary expansion:
\be
\begin{split}
\kappa \reg{\Phi}(z) & = \phi_{(\epsilon)} z^{\epsilon} + c_{22} \phi_{(\epsilon)}^2 z^{2 \epsilon} + c_{33} \phi_{(\epsilon)}^3 z^{3 \epsilon} + \ldots\,, \label{DWPhi} \\
\reg{a}(z) & = \frac{l a_{(0)}}{z} \left[ 1 + b_{11} \phi_{(\epsilon)} z^{\epsilon} + b_{22} \phi_{(\epsilon)}^2 z^{2 \epsilon} + b_{33} \phi_{(\epsilon)}^3 z^{3 \epsilon} + \ldots \right]\,.
\end{split}
\ee
One finds
\be
\begin{split}
c_{22} & = - \frac{\reg{\lambda}_3}{\epsilon \; (d + (v-2) \epsilon)}, \\
c_{33} & = \frac{\reg{\lambda}_3^2}{\epsilon^2 \; (d + (v - 2) \epsilon) (d + (v - 3) \epsilon)} - \frac{\reg{\lambda}_4}{2 \epsilon \; (d + (v - 3) \epsilon)} - \frac{\epsilon \dreg}{4 (\dreg - 1)(d + (v - 3) \epsilon) }\\
b_{11} & = 0, \qquad \qquad b_{22}  = - \frac{1}{4 (\dreg - 1)}, \qquad \qquad b_{33}  = \frac{4 \reg{\lambda}_3}{9 \epsilon \; (\dreg - 1)(d + (v - 2) \epsilon)}.
\end{split}
\ee
Comparison with the results for rigid gravity, \eqref{c22} and \eqref{c33}, shows that $c_{22}$ is unaltered, $c_{33}$ contains the same expressions plus a correction due to the coupling with gravity.  The terms $b_{ii}$ represent their counterparts for the renormalization of the scale factor.  The renormalization of the scalar source, $\phi_{(\epsilon)}$, then proceeds as in the previous section. In particular one defines the regulated source field $\reg{\Psrc}$ and its analog $\reg{\mathfrak{a}}=\exp{\reg{\mathfrak{A}}}$ for the scalar factor.

A crucial step in this process has been the definition of $\reg{\Psrc}$ -- particularly removing the vev terms so that we only work with renormalized sources.  This  regulated source field, $\reg{\Psrc}$, is  uniquely defined for any dimensions $d$ and $\Delta$ (we assume $\Delta > d/2$) by the following conditions:
\begin{enumerate}
\item[(1)] It solves the (regulated) equations of motion.
\item[(2)] It does not depend on boundary coordinates or derivatives with respect to boundary coordinates apart from the implicit, algebraic dependence through $\phi_{(\dreg - \Dreg)}(\bs{x})$.
\item[(3)] It has no vev coefficient, $\phi_{(\Dreg)} = 0$.
\end{enumerate}
Furthermore, the unregulated source field, $\Psrc$, satisfies the same three conditions in the $\epsilon \to 0$ limit. The fact that  condition (3) holds for the unregulated field, \textit{i.e.}, $\phi_{(d)} = 0$, follows from the fact that all terms in the near-boundary expansion of the regulated field $\reg{\Psrc}$ are of the form $z^{n \epsilon}$. Hence, there are no terms $\epsilon$-close to the vev term $z^{d}$ and therefore the coefficient of $z^d$ must remain zero when the $\epsilon \rightarrow 0$ limit is taken.

Conditions (1) -- (3) necessarily imply that both regulated and unregulated source fields represent homogeneous domain wall solutions with vev fixed to zero.  Unlike the bulk field, $\Phi$, dual to the boundary operator $\O$ and obeying Dirichlet boundary conditions, the source field, $\Psrc$, satisfies Neumann conditions at the boundary. Let us stress here that the leading source coefficient $\phi_{(d - \Delta)}$ in $\Phi$ (we assume $\Delta > d/2$) remains identified with the bare UV coupling as dictated by the standard holographic dictionary. Its dependence on the renormalized source, $\fr$, though, is determined by the source field $\Psrc_L$. In this sense, we can think of $\Psrc_L$ as a bulk field dual to the boundary source, $\fr$, in the sense that $\Phi$ is dual to $\O$.

This analysis provides a new understanding of the maximal subtraction scheme defined in \cite{Balasubramanian:2012hb}. In our terminology this scheme satisfies conditions (1) and (3) above. Hence, its zero-momentum limit corresponds to our physical on-shell renormalization scheme. The fact that the maximal subtraction scheme allows for the source field to depend explicitly on boundary momentum means that the dependence between the bare coupling $\phi_{(d-\Delta)}$ and the renormalized coupling $\fr$ becomes non-local. We do not consider this to be a desirable feature, since  textbook QFT renormalization procedures yield a local dependence. 

Let us now analyze the consequences of conditions (1) -- (3). Condition (1) states that the source fields themselves satisfy the equations of motion:
\be
\begin{split}
\partial_r \Psrc & = W'(\Psrc)\,, \label{eqPsi} \\
\partial_r \mathfrak{A} & = - \frac{\kappa^2}{d - 1} W(\Psrc)\,.
\end{split}
\ee
Given $V$, \eqref{VOfW} is a first order differential equation for $W$ and there will be many prepotentials which correspond to a given potential. In \cite{Papadimitriou:2007sj} it was shown that two continuous families of prepotentials, $W^{\pm}_\xi$ exist for a given $V$ at a generic point in the parameter space $(d, \Delta)$. Assuming $\Delta > d/2$ these families can be characterized by the following expansions,
\be
\begin{split}\label{Wpmxi}
W^-_\xi(\Psrc) & = - \frac{d-1}{l \kappa^2} - \frac{d - \Delta}{2 l} \Psrc^2 - \ldots - \frac{\xi}{l \kappa^2} \frac{d-\Delta}{d} (\kappa \Psrc)^{\frac{d}{d - \Delta}} - \ldots\,, \\
W^+_\xi(\Psrc) & = - \frac{d-1}{l \kappa^2} - \frac{\Delta}{2 l} \Psrc^2 - \ldots - \frac{\xi}{l \kappa^2} \frac{\Delta}{d} (\kappa \Psrc)^{\frac{d}{\Delta}} - \ldots\,,
\end{split}
\ee
where $\xi$ is an integration constant. Upon substituting these expressions to \eqref{eqPsi} one finds that the only prepotential satisfying condition (3) is $W^-_0$. Hence, this is the unique prepotential corresponding to the source field associated to the dimensional renormalization procedure.  Notice that this prepotential belongs to the family associated with Neumann boundary conditions imposed on the source field, \cite{Papadimitriou:2007sj}.

This identification allows us to unambiguously relate the field theoretic beta function in the dimensional renormalization scheme with the RG flow given by the homogeneous domain wall solutions. Combining \eqref{betag} with \eqref{eqPsi} we arrive at the result 
\begin{equation} \label{betagDW}
\beta_g(g_L) = \kappa l (W^-_0)' ( \kappa^{-1} g_L )\,.
\end{equation}
With the normalization $l = \kappa = 1$ this gives simply $\beta_g = (W^-_0)'$. This matches the results obtained in the context of holographic cosmology in \cite{Bzowski:2012ih}. Notice, however, that the scheme dictated by holographic dimensional renormalization differs from the usual identification of the \emph{holographic beta function}, $\beta_H$, in \cite{Girardello:1998pd, Freedman:1999gp, deBoer:1999tgo},
\begin{equation} \label{betaH}
\beta_H(g_L) = - \frac{d-1}{\kappa} \frac{W'(\kappa^{-1} g_L)}{W(\kappa^{-1} g_L)}\,.
\end{equation}
This difference arises due to the identification of the renormalization scale $\mu \sim a = \exp(A)$ in \cite{Girardello:1998pd, Freedman:1999gp, deBoer:1999tgo} and an identification $\mu = 1/z$ in this work. These identifications agree at the AdS critical points, as they must, however the usual ambiguity along the flow is resolved with the identification of the source field with the running coupling and its equation of motion \eqref{eqPsi} with the RG equations.

Futhermore, while we have selected the prepotential $W^-_0$ as a necessary consequence of the dimensional renormalization scheme and source redefinition, one is usually supplied with a  \emph{superpotential} for domain walls in supergravity.  In \secref{gppz} we will examine the possibility of taking prepotentials with non-vanishing $\xi$, and note that the dimensional renormalization procedure admits the correct superpotential in the case of relevant deformations.  Then we will compare our result \eqref{betagDW} to the `holographic beta function.' The renormalizability of the QFT and the finiteness of the beta function  can now be stated as the finiteness of the selected prepotential.

\subsection{Example: cubic prepotential}
It is satisfying to demonstrate this procedure and show that it extends to the definition of the beta function to all orders in $\fr$ using the simple solvable example of the cubic prepotential,
\begin{equation}
W(\Phi) = -\frac{d-1}{l \kappa^2} - \frac{w_3}{3 l \kappa^2} (\kappa \Phi)^3\,.
\end{equation}

In the context of relevant deformations the equations of motion \eqref{eqPhi} are usually interpreted as holographic RG equations \cite{deBoer:1999tgo}. For marginal deformations, however, such an interpretation is problematic. Indeed, in this case the system can be solved exactly, and the solution reads
\be
\begin{split} \label{naive_exp}
\kappa \Psrc(z) & = \frac{c_1}{1 - c_1 w_3 \log z}\,,\\ 
\log a(z) & = \log \left( \frac{l c_2}{z} \right) + \frac{c_1^3 w_3}{6(d - 1)} \frac{\log z (c_1 w_3 \log z - 2)}{(c_1 w_3 \log z - 1)^2}\,,
\end{split}
\ee
where $c_1$ and $c_2$ are two integration constants. The interpretation of this solution is problematic as its expansion around $z = 0$ is neither regular nor asymptotically AdS,
\be
\begin{split}
\kappa \Psrc(z) & = c_1 + c_1^2 w_3 \log z + O(\log^2 z)\,, \\
a(z) & = \frac{l c_2}{z} \left(1 - \frac{c_1^3 w_3}{3(d - 1)} \log z + O(\log^2 z) \right).\,
\end{split}
\ee

To fortify the interpretation of \eqref{eqPhi} as RG equations in the marginal case, we need to implement dimensional renormalization, arriving the beta function \eqref{betagDW}.
Starting with the regulated prepotential,
\begin{equation}
\reg{W}(\reg{\Phi}) = -\frac{\dreg-1}{l \kappa^2} - \frac{\epsilon}{2 l} \reg{\Phi}^2 - \frac{\reg{w}_3}{3 l \kappa^2} (\kappa \reg{\Phi})^3\,,
\end{equation}
one can integrate \eqref{eqPsi} to arrive at regulated solutions that depend on the boundary coordinates only through the boundary values $\phi_{(\epsilon)}$ and $a_{(0)}$:
\be
\begin{split}
\kappa \reg{\Psrc}(z) & = \frac{\phi_{(\epsilon)} z^\epsilon}{1 - \phi_{(\epsilon)} \reg{w}_3 \epsilon^{-1} z^\epsilon}\,, \\
\log \reg{\mathfrak{a}}(z) & = \log \left( \frac{l a_{(0)}}{z} \right) - \frac{\epsilon^2}{6 (\dreg - 1) \reg{w}_3^2} \left[ \frac{\epsilon \reg{w}_3 \phi_{(\epsilon)} z^{\epsilon}}{(\epsilon - \reg{w}_3 \phi_{(\epsilon)} z^{\epsilon})^2} + \log \left( 1 - \frac{ \reg{w}_3 \phi_{(\epsilon)} z^{\epsilon}}{\epsilon} \right) \right]\,. \label{regacubic}
\end{split}
\ee
The near-boundary expansion reads
\be
\begin{split}
\kappa \reg{\Psrc}(z) & = \phi_{(\epsilon)} z^\epsilon + \frac{\reg{w}_3 \phi_{(\epsilon)}^2}{\epsilon} z^{2 \epsilon} + \frac{\reg{w}_3^2 \phi_{(\epsilon)}^3}{\epsilon^2} z^{3 \epsilon} + O(z^{4 \epsilon})\,, \\
\reg{\mathfrak{a}}(z) & = \frac{l a_{(0)}}{z} \left[1 - \frac{\phi_{(\epsilon)}^2}{4(\dreg - 1)} z^{2 \epsilon} - \frac{4 \reg{w}_3 \phi_{(\epsilon)}^3}{9 \epsilon(\dreg - 1)} z^{3 \epsilon} + O(z^{4 \epsilon}) \right]\,.
\end{split}
\ee

In order to remove divergences in the scalar sector we trade the bare source, $\phi_{(\epsilon)}$, for the renormalized source $\fr$ via equation \eqref{ren_cond}; this gives
\begin{equation} \label{src_ex}
\fr = L^{-\epsilon} \kappa \reg{\Psrc}(L) = \frac{\phi_{(\epsilon)}}{1 - \reg{w}_3 \epsilon^{-1} L^{\epsilon} \phi_{(\epsilon)}}\,, \qquad\qquad
\phi_{(\epsilon)} = \frac{\fr}{1 + \reg{w}_3 \epsilon^{-1} L^{\epsilon} \fr}\,.
\end{equation}
This cures all the divergences, and the renormalized source field in terms of the dimensionless coupling, $g_L = L^\epsilon \fr$, is
\begin{align} \label{PsiLcubicReg}
\kappa \reg{\Psrc}_L(z) & = \frac{(z/L)^\epsilon g_L}{1 - \reg{w}_3 \epsilon^{-1} ( (z/L)^\epsilon - 1) g_L}\,,
\end{align}
which after sending $\epsilon$ to zero becomes
\begin{equation} \label{PsiLcubic}
\kappa \Psrc_L(z) = \frac{g_L}{1 - w_3 g_L \log (z/L)}\,.
\end{equation}

By comparison with \eqref{naive_exp} we can unambiguously identify $c_1 = g_{L=1}$, the value of the renormalized coupling at the fixed inverse energy scale $L = 1$. Equivalently, $c_1$ can be identified with $g_L$ provided that one substitutes $z$ for $z/L$. It is also important that the $\epsilon \to 0$ divergences in the scale factor \eqref{regacubic} are canceled by the renormalization procedure as well. The renormalized scale factor reads
\begin{equation}
\log \mathfrak{a}_L(z) = \log \frac{l a_{(0)}}{z} - \frac{1}{6(d-1)} \left( \frac{g_L}{1 - w_3 g_L \log (z/L)} \right)^2\,.
\end{equation}
Finally, using \eqref{betadef} the beta function follows from \eqref{src_ex}
\begin{equation}
\beta_g(g_L) = - \epsilon g_L - \reg{w}_3 g_L^2\,.
\end{equation}
Clearly, $\beta_g(g_L) = l \kappa \reg{W}'(\kappa^{-1} g_L)$, in agreement with \eqref{betagDW}. 

\subsection{Correlation functions} \label{correlationfunctions}

In previous sections we concentrated on renormalization of the sources. However, the aim of the renormalization procedure is to make sure that the correlation functions are finite and divergence-free. The one-point function in the presence of sources is affected by the renormalization procedure, since now
\begin{align} \label{ex_1pt}
\< \O(\bs{x}) \>_{ s} & = \lim_{\epsilon \rightarrow 0} \int \D^{\reg{d}} \bs{u} \frac{\delta \reg{S}}{\delta \phi_{(\epsilon)}(\bs{u})} \frac{\delta \phi_{(\epsilon)}(\bs{u})}{\delta \fr(\bs{x})} \nn\\
& = \lim_{\epsilon \rightarrow 0}\int \D^{\reg{d}} \bs{u} \left[ \< \O(\bs{u}) \>_{\text{reg}, s, L} \times \frac{\delta \phi_{(\epsilon)}(\bs{u})}{\delta \fr(\bs{x})} \right] \nn\\
& = - \lim_{\epsilon \rightarrow 0} {2 \Dreg - \dreg\over l \kappa^2} \int \D^{\reg{d}} \bs{u} \left[ \phi_{(\Dreg)}[\phi_{(\epsilon)}(\fr(\bs{u}))] \frac{\delta \phi_{(\epsilon)}(\bs{u})}{\delta \fr(\bs{x})} \right].
\end{align}
The subscript $L$ on the one-point function $\< \O(\bs{u}) \>_{\text{reg}, s, L}$ indicates that it depends on $\fr$ via $\phi_{(\epsilon)}$ as indicated explicitly in the following line.
As an example, consider 2- and 3-point functions of a marginal operator $\O$ in $d=3$ spacetime dimensions on a rigid AdS background with the cubic potential given by \eqref{V}. We will also work in the regularization scheme with $u=2$ and $v=1$, which satisfies the condition $u-v=1$.

For the evaluation of the two-point function, one takes a single functional derivative, with respect to the renormalized source, of the one point function \eqref{ex_1pt} (up to an overall sign.) This gives the 2-point function with sources turned on,
\begin{align} \label{ex_2pt}
\< \O(\bs{x}) \O(\bs{y}) \>_{s} & = {3\over l \kappa^2} \lim_{\epsilon \rightarrow 0} 
\int \D^{\dreg} \bs{u} \left[ \int \D^{\dreg} \bs{v} \left( \frac{\delta \phi_{(\Dreg)}(\bs{u})}{\delta \phi_{(\epsilon)}(\bs{v})} \frac{\delta \phi_{(\epsilon)}(\bs{u})}{\delta \fr(\bs{x})} \frac{\delta \phi_{(\epsilon)}(\bs{v})}{\delta \fr(\bs{y})} \right) \right.\nn\\
& \qquad \left. + \: \phi_{(\Dreg)}(\bs{u}) \frac{\delta^2 \phi_{(\epsilon)}(\bs{u})}{\delta \fr(\bs{x}) \delta \fr(\bs{y})} \right].
\end{align}
The dependence between the bare and renormalized source is given in \eqref{ex1_phi0}, with $g_L = L^\epsilon \fr$, and hence
\begin{equation}
\frac{\delta \phi_{(\epsilon)}(\bs{u})}{\delta \fr(\bs{v})} = \delta(\bs{u} - \bs{v}) \left[ 1 - 2 c_{22} L^\epsilon \fr(\bs{u}) + O(\fr^2) \right]\,.
\end{equation}
For the purposes of the two-point function in the absence of sources, only the leading delta function survives. Evaluating \eqref{ex_2pt} at $\fr=0$ then reduces to finding the free, regulated, momentum-space bulk-to-boundary propagator, $K$, where $\kappa \Phi = K(p,z)\phi_{(\epsilon)} +O(\lambda_3)$, in AdS (see \emph{e.g.} \cite{Skenderis:2002wp})
\begin{align}
 K(z, p) & = z^{\epsilon} e^{-p z}(1 + p z)\,. \label{e:K3}
\end{align}
By dimensional grounds, the integral then selects the coefficient of $z^{\Dreg}$ from $K$, so that in momentum space
\begin{equation} \label{2ptmomentum}
\lla \O(\bs{p}) \O(-\bs{p}) \rra = 3 (l \kappa^2)^{-1} K_{(3 + \epsilon)} = (l \kappa^2)^{-1}p^3\,,
\end{equation}
where the double bracket notation indicated that the overall delta function due to momentum conservation has been dropped.

Clearly, the two-point function did not depend on the source renormalization, however the cubic interaction will have a non-trivial effect on the three-point function. In order to evaluate the three-point function we take another derivative of \eqref{ex_2pt} (with another overall sign.) With sources turned off this gives the three-point function,
\begin{align} \label{3ptspace}
\< \O(\bs{x}) \O(\bs{y}) \O(\bs{z})\> & = - {3\over l \kappa^2} \lim_{\epsilon \rightarrow 0} \left[ \frac{\delta^2 \phi_{(\Dreg)}(\bs{x})}{\delta \phi_{(\epsilon)}(\bs{y}) \delta \phi_{(\epsilon)}(\bs{z})} - 2 c_{22} L^\epsilon \left( \frac{\delta \phi_{(\Dreg)}(\bs{x})}{\delta \phi_{(\epsilon)}(\bs{y})} \delta(\bs{x} - \bs{z}) \right.\right. \nn\\
& \qquad\qquad\qquad \left.\left. + \frac{\delta \phi_{(\Dreg)}(\bs{x})}{\delta \phi_{(\epsilon)}(\bs{y})} \delta(\bs{y} - \bs{z}) + \frac{\delta \phi_{(\Dreg)}(\bs{x})}{\delta \phi_{(\epsilon)}(\bs{z})} \delta(\bs{x} - \bs{y}) \right) \right] \nn\\
& = \lim_{\epsilon \rightarrow 0} \left[ \< \O(\bs{x}) \O(\bs{y}) \O(\bs{z})\>_{\text{reg}} + 2 c_{22} L^\epsilon \left( \< \O(\bs{x}) \O(\bs{y}) \>_{\text{reg}} \delta(\bs{x} - \bs{z}) \right.\right. \nn\\
& \qquad\qquad\qquad \left.\left. + \< \O(\bs{y}) \O(\bs{z}) \>_{\text{reg}} \delta(\bs{y} - \bs{x}) + \< \O(\bs{z}) \O(\bs{x}) \>_{\text{reg}} \delta(\bs{z} - \bs{y}) \right) \right].
\end{align}
The regulated three-point function in momentum space can be evaluated by integrating the product of three bulk-to-boundary propagators. The resulting triple-$K$ integral is divergent and using methods of \cite{Bzowski:2015yxv} one finds
\begin{align}
& \lla O(\bs{p}_1) O(\bs{p}_2) O(\bs{p}_3) \rra_{\text{reg}} = 2 \lambda_3 \int_0^\infty \D z \: z^{-\dreg - 1} K(z, p_1) K(z, p_2) K(z, p_3) \nn\\
& \qquad = \frac{2 \lambda_3}{3 \epsilon} (p_1^3 + p_2^3 + p_3^3) + \text{finite}.
\end{align}
With the value of $c_{22}$ in \eqref{c22} and the result for the two-point function \eqref{2ptmomentum}, we see that the divergence in the regulated three-point function precisely cancels the three contributions of the two-point function in \eqref{3ptspace}. Hence, the renormalized three-point function remains finite.  Note that the renormalized three-point function will contain a scheme-dependent piece dependent on the renormalization scale as well as the $u,v$ parameters as predicted in \secref{genprod}.  For a more thorough discussion of the three-point function, including some examples of explicit scheme-dependence, and the analysis of the four-point function, the interested reader is referred to \cite{Bzowski:2016kni,Bzowski:2015pba}.

\section{Holographic RG flows for relevant deformations}

The dimensional renormalization method has a straightforward extension to the case of relevant deformations. For a single relevant scalar operator the renormalization of the source is unnecessary from the point of view of the UV CFT. In other words, the RG trajectory is well-parameterized by the value of the CFT source, $\phi_{(d - \Delta)}$, at least in the neighborhood of the fixed point. In this case
\begin{equation} \label{trivial_phi}
\phi_{(d - \Delta)} = \varphi = g_L L^{-(d - \Delta)},
\end{equation}
but the \emph{would-be} renormalized source, $\varphi$, does not depend on $L$. Therefore, the beta function for the  source vanishes, $\beta_\varphi = 0$, and the dimensionless coupling has a classical beta function, $\beta_g = - (d - \Delta) g_L$. 
A would-be source field $\Psi_L$ corresponding to \eqref{trivial_phi} is simply $\kappa \Psi_L(z) = z^{d - \Delta} \varphi$. Notice that such a source field solves the Klein-Gordon equation on empty AdS in the absence of any interactions. However, for a non-trivial RG flow this field will not satisfy bulk equations of motion. 

For this reason, when discussing relevant (and irrelevant) deformations, we find it useful to carry out the source redefinition introduced in \secref{genprod}. In this way we consistently work in the on-shell renormalization scheme, where \eqref{ren_cond} is satisfied. The interpretation of the source field $\Psrc_L$ as the running coupling constant follows and we can identify beta functions with those calculated using dimensional regularization  in the field theory.

When we apply holographic dimensional renormalization to a relevant deformation, a subtelty emerges. Recall that for marginal deformations, if the conditions (1) -- (3) of \secref{marginalDW} hold in the regulated theory, then $\phi_{(\Delta)} = 0$ after the regulator is removed. It is well-known that many homogeneous domain wall solutions in supergravity constructions do not satisfy this condition. For a vev coefficient to appear in the $\epsilon \to 0$ limit, the near-boundary expansion of the regulated source field, $\reg{\Psrc}$, must contain a term $\epsilon$-close to the vev term of order $z^{\Dreg}$. One will find such a term, given by:
\begin{equation} \label{vev_from_loc}
\phi_{(\Delta)} = \lim_{\epsilon \rightarrow 0} \phi_{((n-1)(\dreg - \Dreg))}\,,
\end{equation}
where $n$ is an integer defined by
\begin{equation} \label{ncond}
n = \frac{d}{d - \Delta}\,, \qquad\qquad n - 1 = \frac{\Delta}{d - \Delta}\,.
\end{equation}

This will result in a vev coefficent for $\Psi$ provided $\phi_{((n-1)(\dreg - \Dreg))}$ is finite in the $\epsilon\to 0$ limit. Since $\phi_{((n-1)(\dreg - \Dreg))}$ is a local function of the source, so is the vev coefficient and hence the source redefinition remains local.  In the remainder of this section we will use the well-known GPPZ flow \cite{Girardello:1998pd,Girardello:1999bd} as an example to illustrate how dimensional renormalization works in this case.

\subsection{Example: the GPPZ flow} \label{gppz}

The single scalar flow of \cite{Girardello:1999bd} involves an operator $\O$ of dimension $\Delta = 3$ in a $d = 4$ dimensional theory. The bulk supergravity is governed by the superpotential
\begin{equation} \label{Wgppz}
W = -\frac{3}{2 l \kappa^2} \left[ 1 + \cosh \left( \sqrt{\frac{2}{3}} \kappa \Phi \right) \right]\,,
\end{equation}
from which the potential follows
\begin{equation} \label{Vgppz}
V = -\frac{3}{2 l^2 \kappa^2} \cosh^2 \left( \frac{\kappa \Phi}{\sqrt{6}} \right) \left[ 3 + \cosh \left( \sqrt{\frac{2}{3}} \kappa \Phi \right) \right]\,.
\end{equation}
The domain wall solution for \eqref{Wgppz} reads
\begin{align} \label{solgppz}
\kappa \Phi(z) & = \sqrt{6} \arctanh \left( \frac{z  \phi_{(1)}}{\sqrt{6}} \right)\,, \\
a^2(z) & = \frac{1}{z^2} - \frac{ \phi_{(1)}^2}{6}\,.
\end{align}
The radial expansion of the scalar field reads
\begin{equation}
\kappa \Phi(z) = \phi_{(1)} z + \frac{1}{18} \phi_{(1)}^3 z^3 + O(z^5)\,,
\end{equation}
and hence exhibits a non-vanishing vev coefficient, $\phi_{(3)} = \phi_{(1)}^3/18$.

We want to understand this behavior from the point of view of the source redefinition. Our starting point is the potential \eqref{Vgppz}. We expand the potential to the required order and substitute regulated fields and couplings,
\begin{equation} \label{Vgppzreg}
\reg{V} = -\frac{\dreg(\dreg-1)}{2 l^2 \kappa^2} + \frac{\Dreg(\Dreg - \dreg)}{2 l^2} \reg{\Phi}^2 - \frac{\kappa^2}{6 l^2} \reg{\Phi}^4 + \frac{\kappa^2\tilde{\lambda}_4 \epsilon}{4 l^2} \reg{\Phi}^4 + O(\reg{\Phi}^6)\,.
\end{equation}
In general, coefficients in the regulated potential can be $\epsilon$-dependent. In the marginal case, sub-leading terms in the couplings always lead to sub-leading terms in the solutions to the equations of motion and are therefore vanising in the $\epsilon\to 0$ limit. Here, this will not be the case and we keep the sub-leading coupling of order $\epsilon$, denoted $\tilde{\lambda}_4$.

The equations of motion can be solved by expanding fields in the radial variable. The regulated solution, with $\phi_{(\Dreg)} = 0$ reads
\begin{align} \label{Psigppz}
\kappa\reg{\Psi} & = \phi_{(1+\epsilon)} z^{1+\epsilon} + \reg{c}_{33} \phi_{(1+\epsilon)}^3 z^{3+3\epsilon} + O(z^4)\,,
\end{align}
where
\begin{equation}
\reg{c}_{33} = \frac{18 \tilde{\lambda}_4 + (23 - v)}{36(3 - v)} + O(\epsilon)\,.
\end{equation}
The use of the regulated theory is crucial in this statement because we can distinguish a regulated vev term, $\phi_{(3+v\epsilon)} z^{3+v\epsilon}$ from a term local in the bare source, $\phi_{(3+3(u-v)\epsilon)}z^{3+3(u-v)\epsilon}$.  Despite the fact that the constant $\tilde{\lambda}_4$ enters the potential at order $\epsilon$, it does not disappear from the solution when the regulator is removed. This means that  one can obtain a family of scheme-dependent solutions parameterized by $\tilde{\lambda}_4$.  In the regulated theory, all of these solutions have vanishing vev, however in the $\epsilon\to 0$ limit the vev coefficient will be given by \eqref{vev_from_loc}.   The solution corresponding to the GPPZ solution \eqref{solgppz} has
$\tilde{\lambda}_4 = - \frac{17 + v}{18}$.

Stated in terms of the superpotential, this was observed in \cite{Papadimitriou:2007sj}; various homogeneous domain wall solutions parameterized by the value of the vev coefficient $\phi_{(3)}$ can be obtained from choosing different prepotentials $W_\xi^-$. Indeed, when $n$ defined in \eqref{ncond} is an integer, all prepotentials in \eqref{Wpmxi} exhibit regular Taylor expansions. The regulated prepotential $\reg{W}_0^-$ following from the regulated potential \eqref{Vgppzreg} reads
\begin{equation} \label{gppzreg}
\reg{W}_0^- = - \frac{\dreg-1}{\kappa^2 l} - \frac{\dreg-\Dreg}{2 l} \reg{\Phi}^2 - \frac{\kappa^2 \reg{w}_4}{4 l} \reg{\Phi}^4 + O(\reg{\Phi}^5)\,,
\end{equation}
where
\begin{equation}
\reg{w}_4 = \frac{18 \tilde{\lambda}_4 + (23 - v) }{18 (3 - v) } + O(\epsilon)\,.
\end{equation}
Note that for the unregulated prepotential, in the case $n$ is integer, $\xi$ can only be defined relative to a reference prepotential. However, in the regulated case, we consider prepotentials that do not give rise to a regulated vev coefficient which can be unambiguously identified as the family $\reg{W}_0^-$ depending on $\tilde{\lambda}_4$.

With the understanding that the regulated source  only depends locally on the bare source, there is no obstacle to renormalizing the theory in such a way that the source field $\Psrc_L$ remains identified with the running coupling. This can be accomplished using the standard superpotential \eqref{Wgppz}. In order to maintain the on-shell renormalization condition \eqref{ren_cond} we redefine the UV source according to \eqref{src_ren},
\begin{equation}
g_L = \sqrt{6} \arctanh \left( \frac{L  \phi_{(1)}}{\sqrt{6}} \right), \qquad\qquad \phi_{(1)} = \frac{\sqrt{6}}{ L} \tanh \left( \frac{g_L}{\sqrt{6}} \right)\,.
\end{equation}
This gives the running coupling
\begin{equation}\label{psiLgppz}
\kappa \Psrc_L(z) = \sqrt{6} \arctanh \left[ \frac{z}{L} \tanh \left( \frac{ g_L}{\sqrt{6}} \right) \right]\,.
\end{equation}
This expression satisfies the correct normalization conditions,
\begin{align}
\kappa \Psi_L(L) & = g_L\,, \label{first} \\
\lim_{L\to 0} \kappa \Psi_L(z) & = \sqrt{6} \arctanh \left( \frac{z  \phi_{(1)}}{\sqrt{6}} \right)\,. \label{second}
\end{align}
The first equation \eqref{first}, is the on-shell condition stating that the value of the running coupling at $z = L$ equals $g_L$. The second equation \eqref{second} demonstrates that in the UV limit the source field reproduces the  dependence on the bare coupling  expected from the solution \eqref{solgppz}; this is obtained by expanding $g_L=L(\phi_{(1)}+O(L))$, inserting this into \eqref{psiLgppz} and taking $L\to 0$.  

\begin{figure}[ht]
\includegraphics*[width=0.65\textwidth]{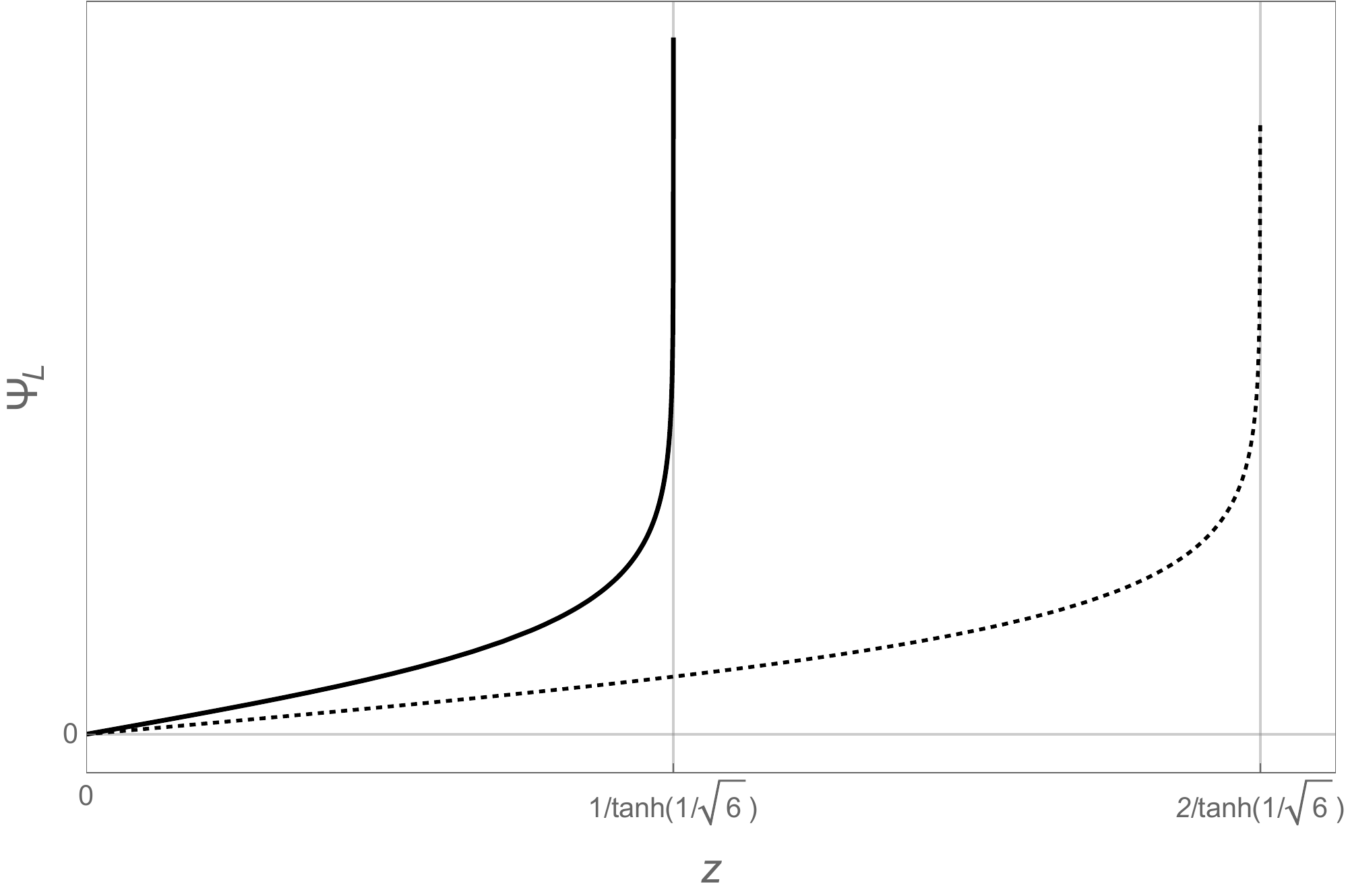}
\centering
\caption{The value of the running coupling $\Psrc_L(z)$ as a function of the inverse energy scale $z$. The two lines represent on-shell renormalization schemes with $\Psrc_L(L) = 1$ with $L = 1$ (solid line) and $L = 2$ (dotted line).\label{fig:gppz}}
\end{figure}
 
\begin{figure}[ht]
\includegraphics*[width=0.46\textwidth]{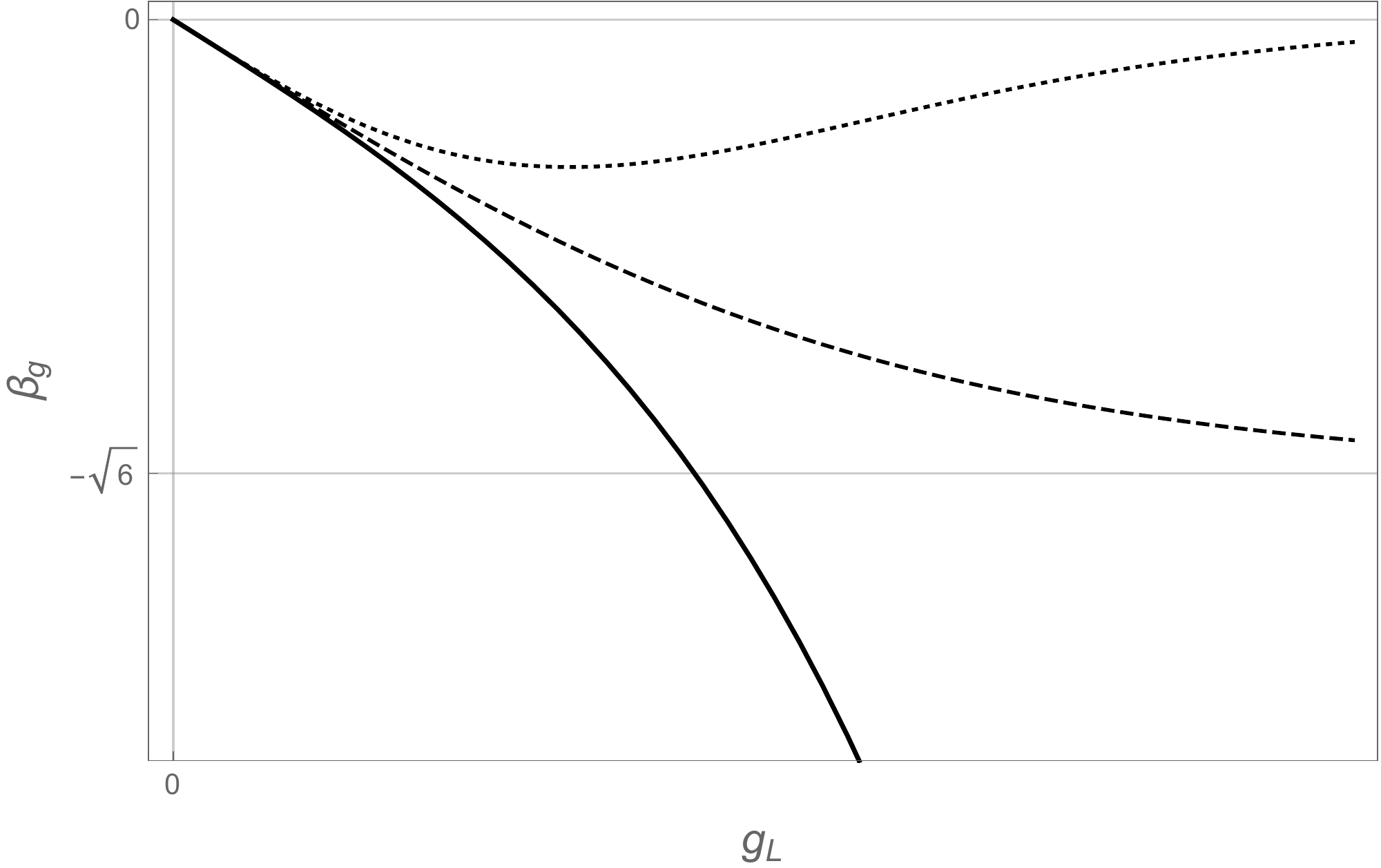}
\qquad
\includegraphics*[width=0.46\textwidth]{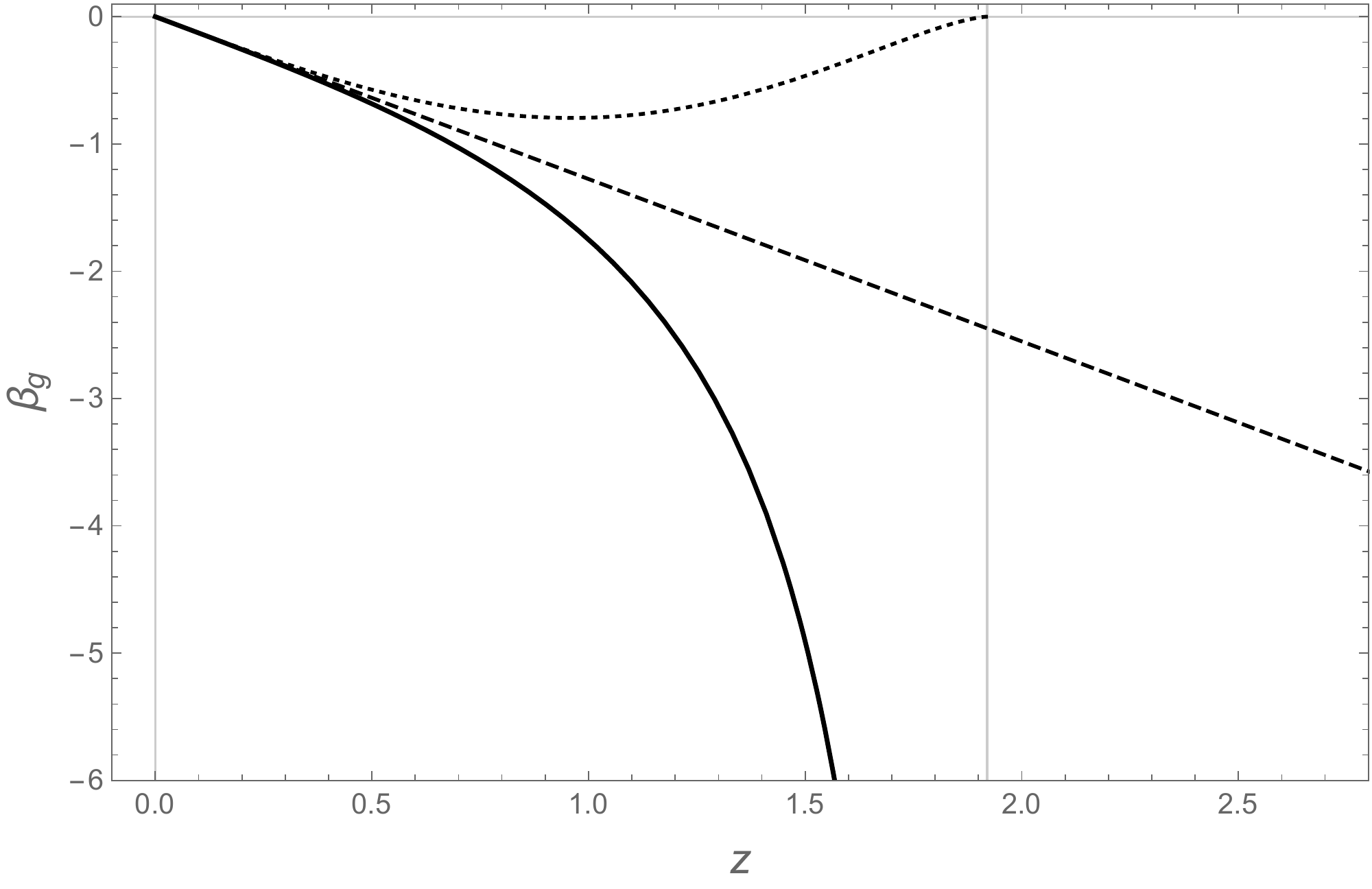}
\centering
\caption{Left: Beta functions as functions of the dimensionless coupling $g_L$. The solid line represents the beta function of the dimensional renormalization scheme, \eqref{betagGPPZ}. The dashed line represents the `holographic beta function' \eqref{betaH} and the dotted line the `proper beta function' \eqref{betaP} as defined in \cite{Anselmi:2000fu}. Right: the same three beta functions, but as a function of the inverse-energy scale $z$. This is obtained by substituting  the solution for the running coupling  \eqref{psiLgppz} with a generic initial condition $g_L = 1$ at $L = 1$ into the beta function.  \label{fig:beta_gppz}}
\end{figure}

In Figure \ref{fig:gppz} we present plots of the running coupling, $\Psrc_L(z)$, in \eqref{psiLgppz} as a function of the inverse energy scale, $z$. The running coupling exhibits a pole at $z=(\tanh( g_L/\sqrt{6}))^{-1}$.  This reflects the fact that the GPPZ flow is singular in the IR.  The position of the singularity can be adjusted using an integration constant to shift $z$.  Here, this freedom is reflected by fixing the value of the source at a different renormalization scale $L$.  The pole in the QFT coupling here is \emph{not} the same as \emph{e.g.} the Landau pole which arises when the running coupling causes perturbation theory to break down.  Rather, the beta function is exact to all orders in the coupling from the QFT perspective, and the diverging coupling reflects the breakdown of the supergravity approximation as the bulk approaches a curvature singularity.

In Figure \ref{fig:beta_gppz} we show the comparison between the beta function of the dimensionally regulated theory,
\begin{equation} \label{betagGPPZ}
\beta_g(g_L) = l \kappa W'(\kappa^{-1} g_L) = - \sqrt{\frac{3}{2}} \sinh \left( \sqrt{\frac{2}{3}} g_L \right)\,,
\end{equation}
represented by the solid line, with other proposals for the beta function. The dashed line represents the `holographic beta function,' \eqref{betaH}, while the dotted line is the `proper beta function,' $\beta_P$, as defined in \cite{Anselmi:2000fu}, up to a factor of $\sqrt{2}$,
\begin{equation} \label{betaP}
\beta_P(g_L) = \beta_H(g_L) \left( -\frac{\kappa^2 l}{d-1} W(\kappa^{-1} g_L) \right)^{-(d-1)/2}\,.
\end{equation}
We have removed a factor of $\sqrt{2}$ from the original definition of \cite{Anselmi:2000fu} in order to match the universal classical CFT scaling behavior, $\beta \sim -g_L$, in the UV. The dimensionally regulated beta function reflects the divergence of the coupling at the location of the bulk singularity.  We do not find this surprising since the presence of the bulk singularity indicates that there is no IR CFT within the supergravity approximation.  Meanwhile, the `proper beta function' indicates the presence of an IR CFT at the location of the singularity, and the holographic beta function gives no indication that it knows about the singularity.

\subsection{Non-perturbative effects emerge} \label{sec:nonpert}

In the previous section we demonstrated how different regularizations of the potential correspond to different prepotentials which imply domain wall solutions with different vevs. The vev can be read from the regulated solution using equation \eqref{vev_from_loc}. This, however, assumes that the local term on its right hand side is finite in the $\epsilon \rightarrow 0$ limit. As we show now, the existence of a finite limit is due to the fact that the quartic term in the potential \eqref{Vgppzreg} takes on a special value, $-\kappa^2/(6 l^2) + O(\epsilon)$. 

Consider a general quartic potential (for simplicity we will take $\lambda_3=0$)
\begin{equation} \label{vquart}
V = -\frac{6}{l^2 \kappa^2} - \frac{3}{2 l^2} \Phi^2 + \frac{\lambda_4}{4 l^2} \kappa^2 \Phi^4 + O(\Phi^5)\,,
\end{equation}
and its regularization
\begin{equation} \label{vgenquart}
\reg{V} = -\frac{\dreg(\dreg-1)}{2 l^2 \kappa^2} + \frac{\Dreg(\Dreg - \dreg)}{2 l^2} \reg{\Phi}^2 + \frac{\lambda_4 + \tilde{\lambda}_4 \epsilon}{4 l^2} \kappa^2 \reg{\Phi}^4 + O(\reg{\Phi}^6)\,,
\end{equation}
with $\Dreg = 3 + v \epsilon$ and $\dreg = 4 + (1 + v) \epsilon$. With general $\lambda_4$ the solution takes form \eqref{Psigppz} with the divergent coefficient
\begin{align} \label{divsol}
\reg{c}_{33} = \frac{2 + 3 \lambda_4}{6 (3 - v) \epsilon} + O(\epsilon^0)\,,
\end{align}
and a finite scheme-dependent contribution. The divergence indicates that no $\epsilon \rightarrow 0$ limit exists and the source redefinition fails. Only when $\lambda_4 = -2/3$, as in \eqref{Vgppzreg}, does the limit exist. Equivalently, the prepotential associated to the general potential \eqref{vgenquart} takes form \eqref{gppzreg} with divergent $\reg{w}_4$,
\begin{equation} \label{w4sol}
\reg{w}_4 = \frac{2 + 3 \lambda_4}{3 (3 - v) \epsilon} + \frac{18 \tilde{\lambda}_4 + (23 - v)}{18 (3 - v)} + O(\epsilon)\,.
\end{equation}
Hence, for general $\lambda_4$, one \emph{must} leave a non-zero $\reg{\xi}$ in the prepotential \eqref{Wpmxi} in order for the $\epsilon \rightarrow 0$ limit of $\reg{W}_{\reg{\xi}}^-$ to exist. To be specific,
\begin{equation} \label{regWeirdW}
\reg{W}_{\reg{\xi}}^- = - \frac{\dreg-1}{\kappa^2 l} - \frac{\dreg-\Dreg}{2 l} \reg{\Phi}^2 - \frac{\reg{w}_4}{4 l \kappa^2} (\kappa \reg{\Phi})^4 - \frac{\reg{\xi}}{\reg{n} l \kappa^2} (\kappa \reg{\Phi})^{\reg{n}} + O(\reg{\Phi}^5)\,,
\end{equation}
with $\reg{\xi} = - \reg{w}_4 + \tilde{\xi} + O(\epsilon)$, where $\tilde{\xi}$ is an arbitrary $\epsilon$-independent constant and $\hat{n}$ is given by \eqref{ncond} defined using $\reg{d}$ and $\reg{\Delta}$. Now the finite $\epsilon \rightarrow 0$ limit exists and the prepotential becomes,
\begin{equation} \label{weirdW}
W_\xi^- = - \frac{3}{l \kappa^2} - \frac{\Phi^2}{2 l} - \frac{\eta}{4 l \kappa^2} (\kappa \Phi)^4 \log (\kappa \Phi) - \frac{\tilde{\eta}}{4 l \kappa^2} (\kappa \Phi)^4 - O(\Phi^5)\,,
\end{equation}
where  $\eta$ is fixed in terms of the divergence of $\reg{w}_4$, and $\tilde{\eta}$ depends on a combination of subleading terms in \eqref{w4sol} and \eqref{regWeirdW},
\begin{equation} \label{eta}
\eta = \lim_{\epsilon \to 0} \epsilon (3 - v) \reg{w}_4 = \frac{2}{3} + \lambda_4\,, \qquad \tilde{\eta} = \tilde{\xi} - \frac{2+3\lambda_4}{12}\,.
\end{equation}

With the prepotential \eqref{regWeirdW}, however, the corresponding domain wall solution \eqref{Psigppz} contains a non-vanishing vev-coefficient,
\begin{align} \label{Psigppz_div}
\kappa\reg{\Phi} & = \phi_{(1+\epsilon)} z^{1+\epsilon} + \reg{c}_{33} \phi_{(1+\epsilon)}^3 z^{3+3\epsilon} + \phi_{(3+v \epsilon)} z^{3 + v \epsilon} + O(z^4)\,,
\end{align}
whose divergence cancels the divergence of the $\phi_{(3+3\epsilon)}$ term 
\begin{equation} \label{phiWithvev}
\phi_{(3+v \epsilon)} = ( -\reg{c}_{33} + O(\epsilon^0) ) \phi_{(1+\epsilon)}^{\frac{3 + v \epsilon}{1 + \epsilon}}\,.
\end{equation}
When the $\epsilon \rightarrow 0$ limit is taken the solution reads
\begin{equation} \label{logsol}
\kappa\Phi = \phi_{(1)} z + \frac{1}{2} \left( \eta \log (z \phi_{(1)}) + ( \tilde{\eta} - \tfrac{1}{4} \eta) \right) \phi_{(1)}^3 z^3 + O(z^4)\,.
\end{equation}

Let us stress that the value of $\eta$ in \eqref{eta} is physical and scheme-independent in the sense that it is uniquely determined by the coefficients of the potential.  In particular the result does not depend on the regularization parameters. The value of $\tilde{\eta}$ in \eqref{eta}, however, is scheme-dependent as it is determined by the specific regularization scheme parameterized by $\tilde{\lambda}_4$ in \eqref{vgenquart} and $\tilde{\xi}$.

Notice  that the general prepotential \eqref{weirdW} with non-zero $\eta$ becomes non-analytic at $\Phi = 0$. Each term of order $\Phi^k$ for $k \geq 4$ is accompanied by $\log^j \Phi$ for $j$ ranging from $0$ to $k-3$. This suggests an emergence of non-local redefinitions of sources and hence non-local beta functions. Indeed, assuming the relation \eqref{betagDW} holds for the prepotential in \eqref{weirdW}, we obtain a beta function with non-analytic logarithmic terms,
\begin{equation}
\beta_g(g_L) = - g_L - g_L^{3} \left( \eta \log g_L + \tilde{\eta} + \frac{\eta}{4} \right) + \ldots\,.
\end{equation}
Integrating the beta function implies the source redefinition,
\begin{equation}  \label{logtermsIng}
\begin{split}
L^{d-\Delta} \phi_{(d - \Delta)}(g_L) &= \exp \left[ - (d - \Delta) \int^{g_L} \frac{\D g'}{\beta_g(g')} \right] \\
&= g_L - \frac{\eta}{2} g_L^3 \log g_L + \frac{\eta - 6 \tilde{\eta}}{12} g_L^3 + O(g_L^4)\,.
\end{split}
\end{equation}
As we can see the source redefinition is non-perturbative, as the logarithm of the renormalized coupling appears. Such a term cannot be seen in perturbation theory around the CFT fixed point.

Dire consequences of logarithmic terms in the prepotential will be analyzed in the following subsection, but let us point out  that, to the best of our knowledge, logarithmic terms in the prepotential are absent in every holographic model descending from supergravity.  In other words, every potential in a supersymmetric theory will have a tuning to the $\lambda_n$ coupling,  analogous to the quartic coupling taking on the value $\lambda_4 = -2/3$ in the GPPZ case, such that the prepotential remains logarithm-free. 

\paragraph{A conjecture:} The fact that this non-analytic, non-perturbative behavior does not occur in any superpotential derived from any theory of supergravity is somewhat suggestive -- the low energy effective theories dictated by a UV complete quantum gravity are all well-behaved QFTs.  We take this as circumstantial evidence to suggest a constraint that any bottom-up holographic model should take into account: in the case that $n$ \eqref{ncond} is integer, the potential must be tuned such that the non-analytic behavior does not arise.  This constraint on the physical couplings in the potential translates into a constraint on the parameter $\eta$;  we conjecture that in any consistent holographic model the parameter $\eta$ in \eqref{weirdW} must be set to zero.  This constraint precludes a term in the beta function proportional to $g_L^n \log g_L$.

Exotic RG flows are studied in \cite{Kiritsis:2016kog} where beta functions are generally non-analytic, containing rational powers of the coupling.  However, in a specific example, \cite{Kiritsis:2016kog} finds that certain non-analytic behavior leads to a multi-valued potential and therefore precludes a unitary holographic realization.  It would be interesting to study these exotic flows using the machinery presented here in an attempt to understand or rule out more general non-analytic behaviors in holographic constructions.

\subsection{Zero-momentum limit and the anomaly}

The source field $\Psrc$ is a zero-momentum solution to the bulk equations of motion. However, it is not clear that $\Psrc$ is the zero-momentum limit of the full bulk solution with the boundary conditions appropriate for the evaluation of correlation functions. In general, there is no reason for this to be the case. A zero-momentum limit of $\Phi$ must satisfy the homogeneous domain wall equations of motion, but there is no guarantee that the corresponding prepotential $W_\xi^-$ has $\xi = 0$.  However, in special cases discussed in \secref{gppz} and \secref{sec:nonpert} we considered general prepotentials, $W_\xi^-$, and hence expression \eqref{logsol} represents the most general zero momentum domain wall solution determined by the prepotential \eqref{weirdW}. In this section we will consider zero-momentum limits that may exhibit non-analytic behavior and therefore we will not work with the renormalized source field.

From the point of view of the dual QFT we should be able to use conformal perturbation theory to express the one-point function in the deformed theory in terms of the UV CFT correlation functions. In momentum space 
\begin{equation} \label{1ptZero}
\< \O(\bs{p}) \> = \sum_{k=0}^{\infty} \frac{(-1)^k \phi_{(d-\Delta)}^k}{k!} \lim_{\bs{p}_j \rightarrow 0} \< \O(\bs{p}) \O(\bs{p}_1) \ldots \O(\bs{p}_k) \>_{\text{CFT}}\,.
\end{equation}
Conservation of momentum implies that $\< \O(\bs{p}) \>$ is proportional to $\delta(\bs{p})$ and hence the one-point function requires evaluation of the total zero-momentum limit $\bs{p}, \bs{p}_j \rightarrow 0$ of all CFT correlation functions on the right hand side. On dimensional grounds,
\begin{equation} \label{zeromom}
\< \O(\bs{p}) \underbrace{\O(0) \ldots \O(0)}_{k} \>_{\text{CFT}} \sim p^{(k+1) \Delta - k d}\,.
\end{equation}
For a relevant scalar ($\Delta < d$) infinitely many terms become IR divergent, a known issue necessitating the use of an IR regulator in massless theories. However, in the context of holography, the existence of the homogeneous domain wall solution implies that the zero-momentum limit exists and is free of an IR regulator. We can examine how holography treats the regulation of IR divergences by taking a zero-momentum limit of the CFT $n$-point function. Practically, this can be obtained by taking (minus) $n-1$ derivatives of the 1-point function and setting $\phi_{(d-\Delta)} = 0$. 

Consider the example analyzed in the previous section. From \eqref{logsol} we find
\begin{equation} \label{logO}
\< \O \> = - \frac{1}{l \kappa^2} \phi_{(1)}^3 \left( \eta \log g + ( \tilde{\eta} - \tfrac{1}{4} \eta) \right),
\end{equation}
where $g = L \phi_{(1)}$ is the dimensionless UV coupling. If we consider the GPPZ theory with $\eta = 0$, then $\< \O \> = -\phi_{(1)}^3\tilde{\eta}/(l \kappa^2) $ and the $4$-point in the UV CFT is constant in momentum space, $\< \O \O \O \O \>_{\text{CFT}} = \frac{6 \tilde{\eta}}{l \kappa^2}$. In position space this corresponds to an ultralocal expression containing a product of delta functions. In the case that $\eta \neq 0$ one encounters a problem: after taking $3$ functional derivative the diverging logarithm precludes the $\phi_{(1)} = 0$ limit.

To understand this situation, recall that the logarithmic term in \eqref{logsol} is non-perturbative: logarithms of the source, $\log \phi_{(1)}$, do not arise in perturbation theory. On the other hand, logarithms of the radial variable, $\log z$, can arise and indicate an anomaly in the UV CFT $n$-point function. We can define the anomaly coefficient $\mathfrak{a}_{\text{UV}}$ for general $n$ \eqref{ncond} via the explicit dependence of the generating function on the renormalization scale,
\begin{equation} \label{defan}
- L \frac{\partial}{\partial L} \mathcal{W}_{\text{CFT}} = \frac{\mathfrak{a}_{\text{UV}}}{l \kappa^2} \int \D^d \bs{x} \, \phi_{(d - \Delta)}^n\,.
\end{equation}
On the level on the $n$-point function this means that
\begin{equation} \label{anom}
-L \frac{\partial}{\partial L} \< \O(\bs{p})\O(\bs{p}_1) \ldots \O(\bs{p}_{n-1}) \>_{\text{CFT}} = (-1)^n n! \mathfrak{a}_{\text{UV}}\,,
\end{equation}
where we dropped the overall delta function due to momentum conservation. 

If $\mathfrak{a}_{\text{UV}}$ vanishes, then the $n$-point function is a constant in momentum space, or  equivalently, an ultra-local product of $(n-1)$ Dirac deltas in position space. In particular, it possesses a zero-momentum limit. If, however, $\mathfrak{a}_{\text{UV}} \neq 0$, then instead of \eqref{zeromom} the $n$-point function behaves logarithmically, $\sim \log p$, preventing an unambiguous zero-momentum limit. From the point of view of the perturbation theory, one would need to introduce an IR regulator in order to analyze this situation. Here, however, holography grants us access to the full IR-complete theory and hence the IR divergences are resolved once the UV theory has been renormalized. The resolution manifests through the non-perturbative appearance of the coupling in \eqref{logO}. This can be regarded as a holographic mechanism of the concept introduced in \cite{Jackiw:1980kv,Appelquist:1981vg}, where it was shown, in general QFTs, that logarithmic IR divergences in correlation functions are regulated by non-perturbative effects that introduce logarithms of the coupling constant, exactly as we  find here.

In the remainder of this section we clarify the intricate UV/IR relation that relates the emergence of the non-perturbative logarithmic $\eta$-term in the prepotential \eqref{weirdW} with the existence of the anomaly, $\mathfrak{a}_{\text{UV}}$, in the UV CFT. We will show that  $\eta$ is in fact proportional to the anomaly coefficient,
\begin{equation} \label{etaOfa}
\eta = - n \mathfrak{a}_{\text{UV}}\,.
\end{equation}
The value of $\eta$ is uniquely determined in terms of the potential of the gravitational theory. On the other hand, the value of the anomaly is determined solely by the UV CFT of the dual theory. Since in the previous section we conjectured that $\eta = 0$ in any consistent holographic theory, this conjecture provides the very strong constraint, $\mathfrak{a}_{\text{UV}} = 0$, on the UV CFT of the field theory dual.

In order to relate $\eta$ and $\mathfrak{a}_{\text{UV}}$ notice that by solving the equation of motion \eqref{eqPsi} for the prepotential \eqref{regWeirdW} we find that the regulated solution exhibits the vev in \eqref{phiWithvev}. For a general value of $n$ this generalizes to,
\begin{equation} \label{special_vev}
\phi_{(\Dreg)} = - \frac{\reg{w}_n}{2 \Dreg - \dreg} \phi_{(\dreg - \Dreg)}^{\reg{n}-1}\,, \qquad\qquad \< \O \> = \frac{\reg{w}_n}{l \kappa^2} \phi_{(\dreg - \Dreg)}^{\reg{n}-1}\,.
\end{equation}
In particular, the bulk field, $\reg{\Phi}$, with the boundary conditions appropriate for the evaluation of correlation functions must satisfy this relation. The divergence of $\reg{w}_n$ propagates to the regulated $n$-point function. Although in the previous section we find a finite vev in the presence of sources, the evaluation of the CFT $n$-point correlation function will not have a well-defined zero-momentum limit.  Since this divergence is due to a divergence in the vev, it must be removed via the addition of a local counterterm, just as in standard holographic renormalization,
\begin{align} \label{special_ct}
S_{\text{ct}} = \left( - \frac{\reg{w}_n}{n} + O(\epsilon^0) \right) \frac{1}{l \kappa^2} \int \D^{\dreg} \bs{x} \sqrt{\gamma_{(0)}} \phi_{(\dreg - \Dreg)}^n L^{((n-1)-v) \epsilon}\,.
\end{align}
The explicit scale-dependence results in the scale-dependent contribution to the CFT generating functional. By taking the derivative with respect to the scale of $\< e^{-S_{\text{ct}}} \> $ and comparing with \eqref{defan} we find
\begin{equation} \label{anomalyeta}
\mathfrak{a}_{\text{UV}} = - \lim_{\epsilon \rightarrow 0} \frac{\epsilon ( (n-1) - v) \reg{w}_n}{n} = - \frac{\eta}{n}\,,
\end{equation}
where we used the generalization of \eqref{eta} to any $n$:
\begin{equation}
\eta = \lim_{\epsilon \rightarrow 0} \epsilon ((n-1) - v) \reg{w}_n\,.
\end{equation}
A condition for $\mathfrak{a}_{\text{UV}}$ to vanish in any field theory dual to a consistent theory of quantum gravity represents a strong testable prediction.

The relation between IR divergences, non-perturbative effects, and the UV anomaly that we have outlined in this section sheds new light on known subtleties in field theory.  For example, there existed a confusing mismatch between one-loop exact beta function and the NSVZ beta function, \cite{Novikov:1983uc,Novikov:1985rd} for the gauge couplings of supersymmetric $SU(N)$ gauge theories in 4 dimensions. This mismatch was explained in \cite{Shifman:1986zi} by careful examination of the IR issues; it was shown that the  difference arises due to the use of the full 1PI coupling in one case and the Wilsonian coupling in the other. The two couplings differ by the inclusion of IR effects, involving a logarithmic redefinition reminiscent of \eqref{logtermsIng}. In the context of this paper, the coupling $\fr$ represents the full 1PI coupling as introduced in \cite{Papadimitriou:2007sj}.  Furthermore, in \cite{ArkaniHamed:1997mj} it is shown that the emergence of such terms is directly related to the existence of the UV axial anomaly, resonating with \eqref{anom}. It would be extremely interesting to see if the analysis leading to \eqref{anomalyeta} can be extended to gauge couplings and other anomalies. 

\subsection{Mixed boundary conditions} \label{sec:further}
In this section we would like to briefly mention the extension of the source renormalization procedure to non-standard boundary conditions.  In this case,
the renormalized coupling will depend non-locally on the bare source. For generic values of $d$ and $\Delta$ the choice of $W_\xi^-$ for $\xi \neq 0$ induces a non-local redefinition of the source via \eqref{special_vev}. The new coupling constant $g_L^{\xi}$ satisfies
\begin{equation} \label{nonlocal_redef}
g_L^\xi = \reg{\Psrc}^\xi(L^\epsilon \phi_{(\epsilon)}) = g_L + \phi_{(\Delta)} L^\Delta + \ldots\,.
\end{equation}
In the spirit of \cite{Papadimitriou:2007sj} we can regard this as imposing `mixed' boundary conditions. Such a redefinition corresponds to a complicated multi-trace deformation. Indeed, the new 1-point function to the leading order is
\begin{align}
\< \O(\bs{x}) \>_\xi & = \frac{\delta S}{\delta g_L^{\xi}(\bs{x})} = \int \D^d \bs{u} \frac{\delta S}{\delta g_L(\bs{u})} \frac{\delta g_L(\bs{u})}{\delta g_L^\xi(\bs{x})} \nn\\
& = \< \O(\bs{x}) \>_0 + \frac{L^\Delta}{2 \Delta - d} \int \D^d \bs{u} \< \O(\bs{u}) \>_0 \< \O(\bs{u}) \O(\bs{x}) \>_0 l \kappa^2+ \ldots\,,
\end{align}
where the subscript $0$ refers to evaluating correlation functions where the source renormalization is local, \emph{i.e.} $\xi=0$. This last term can be regarded as a regulated version of the integral $\int \D^d \bs{u} \< \O^2(\bs{u}) \O(\bs{x}) \>_0$, a sign of the double-trace deformation. This behavior is similar to that encountered in the analysis of $T \bar{T}$-type deformations, \cite{McGough:2016lol,Bzowski:2018pcy}, where the new source is a linear combination of the old source and vev.  A careful analysis of this type of source redefinition, which is unusual from the QFT perspective but of obvious interest in holographic constructions, is a promising topic for future work.

\section{Deformation by an irrelevant operator}

Finally, we would like to demonstrate the application of the dimensional renormalization procedure in the case that an IR CFT is deformed by an irrelevant operator.  As an illustrative example, we will consider the flow in AdS$_3$  between two $d=2$ CFTs \cite{Berg:2001ty}.  The identification of renormalization scheme for this model should provide the precision necessary to make checks between the gravity side and the recently proposed field theory dual \cite{Tong:2014yna}. Specifically, the expansion of this work to renormalized stress tensor correlation functions could be used to match $c$-funcitons computed directly in the CFT.

We begin by examining the superpotential and potential for the truncation of three-dimensional supergravity to a single active scalar. Restoring factors of $\kappa$ and the AdS radius, and using a canonically normalized scalar field, the expressions of \cite{Berg:2001ty} become:
\be 
\begin{split}
W&=-{13+20\cosh(\kappa\Phi )-\cosh(2\kappa\Phi) \over 32 l_{UV} \kappa^2} \\
& = -{1\over l_{UV} \kappa^2} -{\Phi^2 \over 4 l_{UV}} +O(\Phi^4)  \\ 
& = -{1\over l_{IR} \kappa^2} +{3 \over 4 l_{IR}}\(\Phi - \Phi_*\)^2 +O((\Phi-\Phi_*)^3)\,.
\end{split}
\ee
The flow is between the critical points of the potential located at $\Phi=0$ (UV) and $\kappa\Phi_*=\arcosh(5)$ (IR). The two AdS radia are related via $l_{UV}=2l_{IR}$.  From the expansion of the potenial,
\be
\begin{split}
V&={(3+\cosh(\kappa \Phi))^2\(-21-12\cosh(\kappa \Phi)+\cosh(2\kappa \Phi)\)\over 512 l_{UV}^2\kappa^2}\\
&= -{1\over l_{UV}^2\kappa^2} - {3\over 8l_{UV}}\Phi^2 + O(\Phi^4) \\
&= -{1\over l_{IR}^2\kappa^2} + {21\over 8l_{IR}}(\Phi-\Phi_{*})^2 +O((\Phi-\Phi_{*})^3) \,,
\end{split}
\ee
we see that the UV and IR masses correspond to a deformation in the UV with $\Delta^{UV} =3/2$ that flows to an IR CFT with an irrelevant deformation, $\Delta^{IR}=7/2$.

This model represents a very rare case where the domain wall between two regular fixed points is known analytically.  Integrating the equations of motions we find:
\be \label{BergSamtSoln}
{ (5-\cosh(\kappa\Phi))(\cosh(\kappa\Phi)+1)^2 \over (\cosh(\kappa\Phi) -1)^3} = {128\e^{3r/l_{UV}} \over \phi_{(1/2)}^6} \,, \quad \e^{6A(r)} = {c_2(5-\cosh(\kappa\Phi))^4 \over (\cosh(\kappa\Phi)+1)(\cosh(\kappa\Phi)-1)^6}\,,
\ee
where one integration constant has been fixed in terms of the UV source, $\phi_{(1/2)}$, and the constant $c_2$ can be absorbed by a shift in $r$.
Expanding around the UV fixed point in powers of $z_{UV}=l_{UV}{\rm e}^{-r/l_{UV}}$, we notice that the situation is the same as the \secref{gppz}; there appears to be a non-vanishing vev coefficient, $\phi_{(3/2)}$,
\be 
\kappa \Phi = \phi_{(1/2)} \sqrt{z_{UV}} + {1\over 48}\phi^3_{(1/2)}\( z_{UV}\)^{3/2} +O(z_{UV}^{5/2})\,.
\ee
This occurs because the condition \eqref{ncond} is satisfied with $n =4$. Going through the source renormalization procedure for a general potential with the expansion \eqref{vgenquart} we find that the expansion coefficient,
\be
\phi_{(3/2 +\epsilon)}= {4 \lambda_4  + 1 \over 4(3-v)\epsilon}  \phi_{(1/2)}^3+ O(\epsilon^0)\,,
\ee
has a finite $\epsilon \to 0$ limit due to a cancellation with a special value of the quartic coupling, $\lambda_4 = -1/4$.  This is the same type of cancellation, which occurs in all supergravity domain walls, that was noted in \secref{sec:nonpert} and motivates the conjecture at the end of that section.  

One can also expand around the IR fixed point in powers of $z_{UV}^{-1}$,
\be 
\kappa \Phi = {\arcosh(5)} - \sqrt{\frac23}{512\over9\phi_{(1/2)}^6}z_{UV}^{-3}+O(z_{UV}^{-6})\,.
\ee
However, this needs to be modified if we want to read off the CFT data as usual from a near-boundary expansion.  First, we should shift the field so that the IR critical point is the origin of field space: $\kappa\Phi = \kappa\tilde{\Phi} + \arcosh(5)$.  Second, we should rescale the radial coordinate so that the radial expansion is given in terms of the radius of the IR AdS: $z_{IR} = {\rm e}^{-r/l_{IR}} = {\rm e}^{-2 r/l_{UV}}=z_{UV}^2 $.  Finally, we can write the solution in terms of the bare IR source, $\tilde{\phi}_{(-3/2)}$. Then, the IR expansion, at $r=-\infty$, is given in terms of $z_{IR}^{-1}$ by,
\be 
\kappa \tilde{\Phi} = \tilde{\phi}_{(-3/2)}z_{IR}^{-3/2}+\sqrt{\frac32}{5 \over 4}\tilde{\phi}_{(-3/2)}^2z_{IR}^{-3}+O(z_{IR}^{-9/2})\,.
\ee
Thus, we see that the flow into the IR fixed point matches the expectations for a deformation by an irrelevant operator with weight $\Delta^{IR}=7/2$ and exhibits no vev. Furthermore, the condition \eqref{ncond} is not satisfied for integer $n$ in the IR, so the renormalization procedure is straightforward and no additional counterterms are needed.

\begin{figure}[ht]
\includegraphics[width=0.65\textwidth]{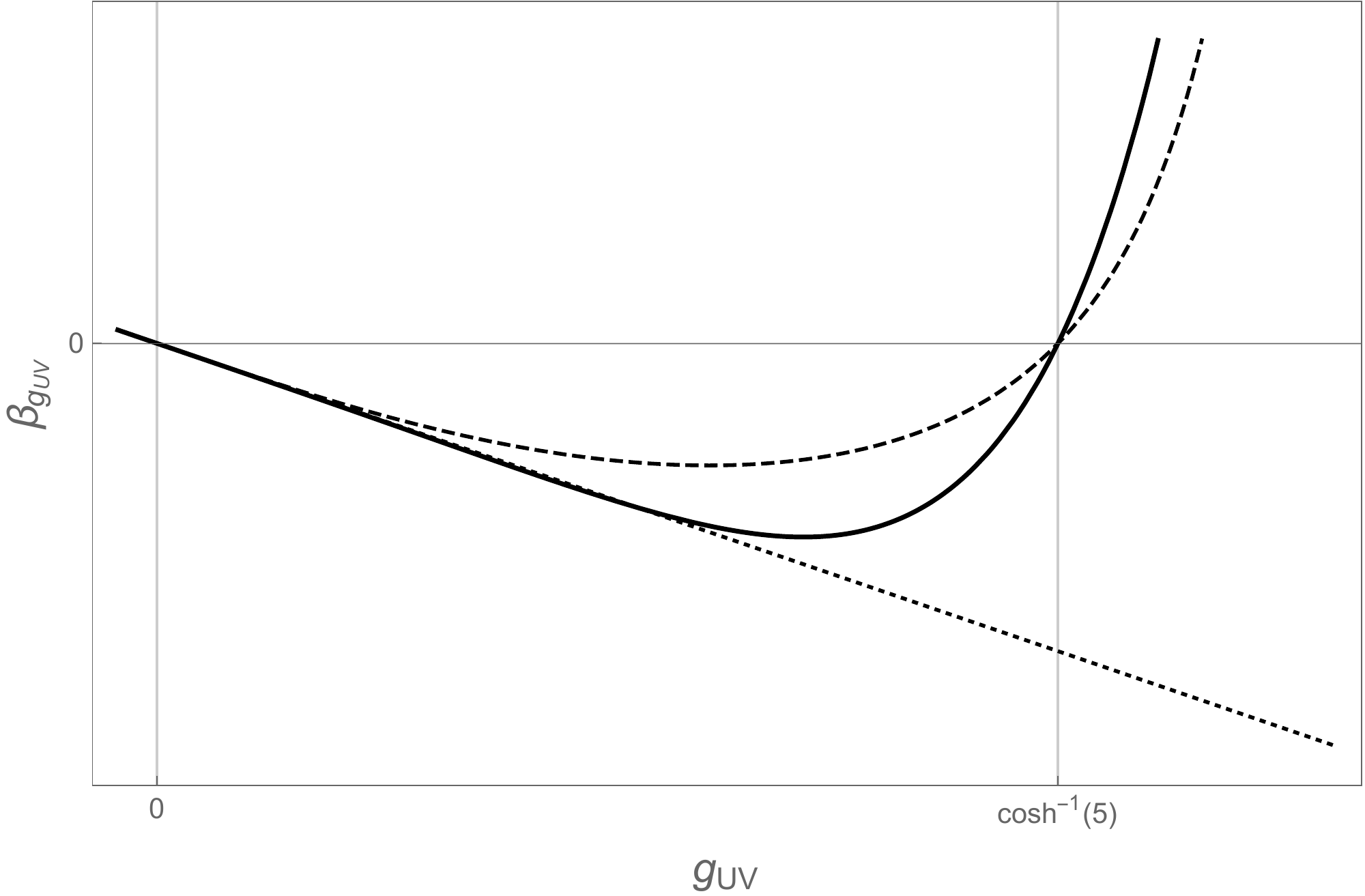}
\centering
\caption{Beta functions for the UV coupling in the Berg-Samtleben flow. The solid line represents the beta function of the dimensional regularization, the dashed line `holographic beta function' \eqref{betaH}, and the dotted line the classical beta function of the UV coupling as advocated in \cite{Berg:2002hy}.\label{fig:bs}}
\end{figure}

The beta function given by \eqref{betagDW} is analytic and non-perturbative.  The  exact expression given by solving the cubic equation \eqref{BergSamtSoln} is not enlightening, but it is plotted in Figure \ref{fig:bs}.  However, we can expand the beta function around both the UV and IR fixed point:
\be
\begin{split}
\beta_{g_{\text{UV}}}(g_{\text{UV}}) &= l_{\text{UV}} \kappa W'(\kappa^{-1} g_{\text{UV}}) = - \frac{1}{2} g_{\text{UV}} - \frac{g_{\text{UV}}^3}{48} + O(g_{\text{UV}}^5)\,, \\
\beta_{g_{\text{IR}}}(g_{\text{IR}}) &= l_{\text{IR}} \kappa W'(\kappa^{-1} g_{\text{IR}}) = \frac{3}{2} g_{\text{IR}} + \frac{15}{8} \sqrt{\frac{3}{2}} g_{\text{IR}}^2 + O(g_{\text{IR}}^3)\,.
\end{split}
\ee
First, we note that the UV expansion does not agree with the result of \cite{Berg:2002hy}; their result corresponds to taking only the first term in the UV expansion.   The reason for this is that the beta function is scheme-dependent and while the authors carry out traditional holographic renormalization and arrive at the classical beta function for the bare source, we have computed the beta function for the  renormalized source, corresponding to  dimensional renormalization on the field theory side.  Second, we note that this allows us to define the beta function to all orders in perturbation theory which can also be analyzed from an IR perspective.

\section{Conclusions}
In this paper we have presented the first instance of a bulk holographic renomalization scheme which corresponds to a know field theory renormalization scheme: dimensional renormalization.  In this scheme we can identify the running coupling as a local function of the bare source which satisfies the bulk equations of motion in the zero-momentum limit, or equivalently, the RG equations.  Using this identification, the holographic beta function is given by $W'$ as opposed to $ -(d-1) W'/W$, as previously proposed.  Furthermore, the prepotential is uniquely determined in the dimensional regularization scheme as the limit of the specific family of regulated prepotentials, $\reg{W}^-_0$, which produce regulated solutions with vanishing vev-coefficients, $\phi_{(\Dreg)} = 0$.

The process of source renormalization in the holographic renormalization scheme presented here allows us to understand the deformation of CFTs by marginal operators which spoil asymptotically AdS boundary conditions.  Despite the emergence of an infinite tower of logarithmic terms, we identify the renormalized source uniquely. Additionally, we establish the relation between the holographic renormalization scheme as the on-shell renormalization condition of the dual QFT, for the first time making a direct connection to the well-known renormalization scheme of the dual QFT. This will be used to better understand the holographic correspondence for marginal deformations such as the confining field theory dual to the Klebanov-Strassler throat.  Furthermore, the process can be extended to relevant and irrelevant deformations, offering new precision in many known holographic scenarios.

The study of well-known supergravity domain walls using this technique indicates that specific tunings in supergravity potentials prevent a non-analytic behavior of the beta function which we relate to the absence of a conformal anomaly.  We conjecture that this cancellation in necessary and should be engineered into bottom-up holographic constructions.  This conjecture can be related to a statement that the conformal anomaly should vanish in UV holographic CFTs.

\section*{Acknowledgments} We would like to thank  Guido Festuccia, Vladimir Prochazka and Kostas Skenderis for useful discussions. This work is supported by the Swedish Research Council (VR) and the Knut and Alice Wallenberg Foundation under grant 113410212.


\providecommand{\href}[2]{#2}\begingroup\raggedright\endgroup

\end{document}